\documentclass[paper, 12pt, letterpaper]{JHEP}
\usepackage{amsmath}
\usepackage{amssymb}
\usepackage{amsfonts}
\usepackage{mathrsfs}
\usepackage{bbm}
\usepackage{bm}

\usepackage{epsfig,rotating}

\usepackage{longtable,booktabs}

\def\Bbb{\mathbb} \def\C{{\Bbb C}} \def\R{{\Bbb R}} \def\Z{{\Bbb Z}}
  \def\L{{\Bbb L}} \def\P{{\Bbb P}}

\def\p{{\bf p}}

\def\vint{{\int \hspace{-0.47cm} \int}}

\def\Der{{\rm Der}}

\def\N{\Bbb{N}}

\def\ev{{\rm ev}}
\def\red{{\rm red}}
\def\res{{\rm res}}
\def\hcA{{\hat {\cal A}}}

\def\A{\Bbb{A}}

\def\Hom{\operatorname{Hom}} 
 \def\Spec{\operatorname{Spec}}
  \def\vol {{\rm
    vol}}

\def\cA{{\cal A}}

\def\cO{{\cal O}}

\def\cT{{\cal T}}

\def\cC{{\cal C}}

\def\tr{{\rm tr}}
\def\str{{\rm str}}
\def\id{{\rm id}}
\def\ker{{\rm ker}}  
\def\im{{\rm im\,}}  
  \def\dim{{\rm dim}}
  \def\d{{\rm d }}
     \def\det{{\rm det}}
  
\def\End{{\rm End}}  
\def\Spec{{\rm Spec}}

\def\2Cat{{\rm 2Cat}}

 \def\cN{{\cal N}}     
\def\cE{{\cal E}}
\def\ucO {\underline \cO}\def\ucC {\underline \cC}

\font\gothics=ygoth at 10pt 
\def\gm{\hbox{\gothics m}}

\newcommand{\lsc}{\{\hspace{-0.1cm}[}
\newcommand{\rsc}{]\hspace{-0.1cm}\}}
\newcommand{\lbr}{(\hspace{-0.1cm}(}
\newcommand{\rbr}{)\hspace{-0.1cm})}

\newcommand{\beq}{\begin{equation}} \newcommand{\eeq}{\end{equation}}
\newcommand{\bea}{\begin{eqnarray}} \newcommand{\eea}{\end{eqnarray}}
\newcommand{\beann}{\begin{eqnarray*}}
  \newcommand{\eeann}{\end{eqnarray*}}
\newcommand{\bfig}{\begin{figure}} \newcommand{\efig}{\end{figure}}
\newcommand{\nn}{\nonumber}
\newcommand{\ba}{\begin{array}}\newcommand{\ea}{\end{array}}
\newcommand{\CC}{\mathcal{C}}
\newcommand{\CD}{\mathcal{D}}
\newcommand{\CE}{\mathcal{E}}

\newcommand{\FC}{\mathbbm{C}}                           
\newcommand{\FR}{\mathbbm{R}}                           
\newcommand{\CO}{\mathcal{O}}
\newcommand{\di}{\mathrm{i}}                            
\newcommand{\unit}{\mathbbm{1}}                         
\newcommand{\sEnd}{\mathrm{End}\,}

\newcommand{\dpar}{\partial}                           
\newcommand{\der}[1]{\frac{\dpar}{\dpar #1}}
\newcommand{\remark}[1]{}
\newcommand{\bz}{{\bar{z}}}
\newcommand{\bZ}{{\bar{Z}}}
\newcommand{\embd}{{\hookrightarrow}}

\newcommand{\derr}[2]{\frac{\dpar #1}{\dpar #2}}        

\newtheorem{Proposition}{Proposition}[section]
\newtheorem{Definition}{Definition}[section]
\newtheorem{Theorem}{Theorem}[section]
\newtheorem{Lemma}{Lemma}[section]
\newtheorem{Corrolary}{Corrolary}[section]

\newcommand{\be}{\begin{equation}} \newcommand{\ee}{\end{equation}}

\newcommand{\bp}{\begin{Proposition}}  \newcommand{\ep}{\end{Proposition}}
\newcommand{\bt}{\begin{Theorem}} \newcommand{\et}{\end{Theorem}}
\newcommand{\bl}{\begin{Lemma}} \newcommand{\el}{\end{Lemma}}
\newcommand{\bc}{\begin{Corrolary}} \newcommand{\ec}{\end{Corrolary}}
\newcommand{\bd}{\begin{Definition}} \newcommand{\ed}{\end{Definition}}

\title{Generalized Berezin-Toeplitz quantization of K{\"a}hler supermanifolds}

\author{~~~Calin Iuliu-Lazaroiu, Daniel McNamee and Christian S{\"a}mann
\\Trinity College Dublin\\
Dublin 2, Ireland\\
calin,~danmc,~saemann@maths.tcd.ie}

\abstract{We extend the construction of generalized Berezin and
Berezin-Toeplitz quantization to the case of compact Hodge
supermanifolds. Our approach is based on certain super-analogues of
Rawnsley's coherent states. As applications, we discuss the quantization of affine and projective superspaces. 
Furthermore, we propose a definition of supersymmetric sigma-models
on quantized Hodge supermanifolds. The corresponding quantum field theories
are finite and thus yield supersymmetry-preserving regularizations
for QFTs defined on flat superspace. }

\preprint{TCDMATH 08-17}


\begin{document}

\section{Motivation and context}

Geometric quantization \cite{Woodhouse:1992de} and its variants (such as
Berezin and Berezin-Toeplitz quantization) arose as techniques for
`quantizing' symplectic manifolds in an attempt to give rigorous
formulations to the canonical quantization procedure for systems with a
finite number of degrees of freedom.  Beyond its importance in physics, this
theory leads to certain notions of `quantum' symplectic geometry, which are of
independent mathematical interest. It also leads to a certain class of
regularizations for quantum field theories \cite{Grosse:1995ar}. Finally, the
worldvolumes of D-branes placed in certain superstring vacua can be described
in terms of quantized spaces \cite{Myers:1999ps, Seiberg:1999vs}.

Since many classical mechanical models admit canonical formulations containing
both even and odd variables, it is natural to extend such quantization
prescriptions to the case of symplectic supermanifolds.  In particular,
quantized Hodge supermanifolds should provide
supersymmetry-preserving regularizations of supersymmetric quantum field
theories.

Geometric quantization was independently developed by Kostant
\cite{Kostant-1970aa,Kostant:1975qe} and Souriau \cite{Souriau-1970aa} and
proceeds in two steps. First, one fixes a positive complex line bundle over
the manifold to be quantized, whose space of $L^2$-sections yields a {\em prequantization}. Second, one
picks a polarization on this bundle, which can be used in order to reduce the
space of $L^2$-sections to a proper subspace, thereafter identified with the
quantum Hilbert space. The question of prequantization for supermanifolds was
considered in \cite{Kostant:1975qe}, later expanded on in
\cite{Tuynman:1992zm}. The appropriate definition of positive super line
bundles was given in \cite{MR1032867}. 

In the present paper, we do not consider geometric quantization but the
closely related Berezin and Berezin-Toeplitz methods, extending them to
Hodge supermanifolds. More precisely, we consider the superextension of the generalized
Berezin and Berezin-Toeplitz quantizations of \cite{IuliuLazaroiu:2008pk},
which subsume the classical Berezin and Berezin-Toeplitz cases. 
Some previous work in this direction, though restricted to the case of certain
homogeneous supermanifolds, can be found\footnote{The work cited relies
on group theoretic methods and on a super-generalization of Perelomov's
coherent states \cite{Perelomov:1986tf}, none
of which can can be applied directly to general Hodge supermanifolds as
defined below.} in \cite{ElGradechi:1993gq,Grosse:1995pr,Murray:2006pi}.

Berezin quantization was introduced in
\cite{Berezin:1974du} while its Berezin-Toeplitz variant
is discussed e.g.\ in \cite{Tuynman:1987jc} and more recently in \cite{MaMarinescu}. Generalized Berezin and Berezin-Toeplitz
quantizations were defined and analyzed in \cite{IuliuLazaroiu:2008pk} and
include interesting new possibilities, such as {\em Berezin-Bergman
quantization}.  The latter prescription is natural
in contexts arising from algebraic geometry. In the present paper, we extend the results of
\cite{IuliuLazaroiu:2008pk} to the class of Hodge supermanifolds.

The paper is organized as follows. Section 2 recalls some basic notions of
supergeometry and introduces the concept of Hodge supermanifolds.  In Section
3, we give the construction of supercoherent states and of generalized Berezin and
Berezin-Toeplitz quantizations for Hodge supermanifolds, as well as a brief
discussion of their properties. Section 4 considers various special cases: the
analogues of classical Berezin and Berezin-Toeplitz quantizations, the
quantizations of affine and projective superspaces as well as Berezin-Bergman
superquantization. The last section shows how one can employ our methods 
to construct supersymmetry-preserving regularizations of supersymmetric
quantum field theories.

\section{Hodge supermanifolds, polarizations and Bergman supermetrics}

In this section, we recall some basic notions from the theory of
supermanifolds. The reader can consult \cite{Manin:1988ds,Varadarajan:2004yz}
for further details. To understand some of the concepts presented in the following, the reader might find it helpful to first study the corresponding definitions for ordinary manifolds as given in \cite{IuliuLazaroiu:2008pk}.

\subsection{Super Hermitian pairings}

Recall that a complex supervector space $E$ is a vector space over the complex
numbers endowed with a $\Z_2$-grading $E=E_+\oplus E_-$. 
A {\em super
Hermitian pairing} on $E$ is a $\C$-sesquilinear even form $(~,~):E\times
E\rightarrow \C$ which is graded-Hermitian, i.e.\ it satisfies the condition
\beq
\label{superherm}
(s,t)=(-1)^{\tilde{s}\tilde{t}}\overline{(t,s)} \eeq for any two
$\Z_2$-homogeneous elements $s,t$ of $E$ with degrees
$\tilde{s},\tilde{t}$. Our convention for sesquilinear forms is that they are
antilinear in the {\em first} variable. 

Evenness of the pairing implies that $(s,t)$ vanishes unless
$\tilde{s}=\tilde{t}$. Hence a super Hermitian pairing is completely
determined by its restrictions to $E_+$ and $E_-$. Relation (\ref{superherm})
shows that the first restriction is a Hermitian form on $E_+$, while the
second is anti-Hermitian on $E_-$. Thus a super
Hermitian form can be expressed as: \beq (s,t)=(s_+,t_+)_++i(s_-,t_-)_-
\eeq where $(~,~)_\pm$ are Hermitian pairings on $E_\pm$. Here $s=s_++s_-$ and
$t=t_++t_-$ are the decompositions of $s,t$ into even and odd
components. Conversely, a choice of Hermitian forms on $E_\pm$ determines a
super Hermitian pairing on $E$.

A super Hermitian pairing $(~,~)$ is called {\em nondegenerate} if it is
nondegenerate as a sesquilinear form, i.e.\ if vanishing of
$(s,t)$ for all $t$ implies $s=0$. This amounts to the requirement that
$(~,~)_+$ and $(~,~)_-$ are both nondegenerate. A super Hermitian pairing is called a {\em
superscalar product} if it is nondegenerate and if $(~,~)_+$ is
positive-definite on $E_+$. 

A super Hermitian form on $E$ induces an even antilinear map
  $\Phi:E\rightarrow E^{\rm *}=\underline{\Hom}_{\C} (E,\C)$ given by: \beq \nn
  \Phi(s)(t)=(s,t)~~, \eeq which is bijective iff the pairing is
  nondegenerate. In that case, the dual supervector space $E^*$ has an induced super
  Hermitian pairing $(~,~)_*$ given by
\beq
  (\eta,\rho)_*=(-1)^{\tilde{\eta}\tilde{\rho}}(\Phi^{-1}(\rho),\Phi^{-1}(\eta))~~.
  \eeq 

Let us fix a nondegenerate super Hermitian pairing on $E$. The {\em super
Hermitian conjugate} of a homogeneous linear operator $A$ on $E$ is defined
through: \beq (As,t)=(-1)^{\tilde{A}\tilde{s}}(s,A^\dagger t)~~~\forall s,t\in
E ~~{\rm homogeneous}~~ \eeq and extended to inhomogeneous operators in the
obvious manner.  When $A$ is even ($A=A_++ A_-$ with $A_\pm\in
\End(E_\pm)$), this boils down to $A^\dagger =A_+^\dagger \oplus
A_-^\dagger$, where $A_\pm^\dagger: E_\pm\rightarrow E_\pm$ are the Hermitian
conjugates of $A_\pm$ with respect to $(~,~)_\pm$. When $A$ is odd ($A=A_1+ A_2$ with $A_1:E_-\rightarrow E_+$
and $A_2:E_+\rightarrow E_-$), we find $A^\dagger=i(A_2^\dagger +
A_1^\dagger)$ i.e.\ $(A^\dagger)_1=iA_2^\dagger$ and
$(A^\dagger)_2=iA_1^\dagger$, where $A_1^\dagger:E_+\rightarrow E_-$ and
$A_2^\dagger:E_-\rightarrow E_+$ are
the Hermitian conjugates of $A_1$ and $A_2$ with respect to the pairings
$(~,~)_+$ and $(~,~)_-$ on $E_+$ and $E_-$.

Super Hermitian conjugation gives a conjugation of the superalgebra
$({\underline \End}(E),\circ)$, i.e.\ an even and involutive antilinear antiautomorphism of
this superalgebra. In particular, we have: 
\beq
(AB)^\dagger=(-1)^{\tilde{A}\tilde{B}}B^\dagger A^\dagger~~. 
\eeq 
This superalgebra is also endowed with the usual
supertrace $\str:{\underline \End}(E)\rightarrow \C$, which is an even map and satisfies: \beq
\str(AB)=(-1)^{\tilde{A}\tilde{B}}\str(BA)~~.  \eeq Notice\footnote{It
suffices to check this for even operators $A$ since $\str(A)$ vanishes when
$A$ is odd.} that $\str(A^\dagger)=\overline{\str(A)}$.

The underlying supervector space ${\underline \End}(E)$ carries the {\em super Hilbert-Schmidt
pairing} induced by $(~,~)$, which is defined through: \beq \langle A,B\rangle_{HS}=\str(A^\dagger
B)\in \C~~.  \eeq This is itself a non-degenerate super Hermitian pairing on ${\underline
  \End}(E)$, and in particular it satisfies: 
\beq 
\overline{\langle
A,B\rangle_{HS}}=(-1)^{\tilde{A}\tilde{B}} \langle B, A\rangle_{HS}~~.
\eeq
Notice that $\langle ~,~\rangle_{HS}$ need not be a superscalar product even when $(~,~)$ is.

\subsection{Supermanifolds}

Throughout this paper we will work with supermanifolds in the sense of
Berezin (see \cite{Manin:1988ds}). Recall that
a {\em superspace} over $\C$ is a locally super ringed space over the complex
numbers, i.e.\ a pair $(X,\cA)$ where $X$ is a topological space and $\cA$ is a
sheaf of superalgebras over $\C$ such that the stalk $\cA_x$ of $\cA$ at any
point $x\in X$ is a local superalgebra. Given a superspace, we let
$\cA_n\subset \cA$ be the subsheaf of nilpotent elements of $\cA$, and set
$\cA_\red:=\cA/\cA_n$ and $\hcA=\cA_n/\cA_n^2$. We say that a superspace
$(X,\cA)$ is a real (resp.\ complex) {\em supermanifold} of dimension $(m|n)$ if:

\begin{enumerate}
\item[(1)] $(X,\cA_\red)$ is the locally ringed space of smooth (resp. holomorphic)
complex-valued functions associated with a real (resp. complex) manifold structure on $X$ of
real (resp. complex) dimension $m$,
\item[(2)] $\hcA$ is locally free of purely
odd finite rank $0|n$ as a sheaf of $\cA_\red$-supermodules,
\item[(3)] $\cA$ and $\wedge^n_{\cA_\red}\hcA $ are {\em locally} isomorphic as
sheaves of superalgebras over $\cA_\red$.
\end{enumerate}

The natural surjection $\cA\rightarrow \cA_\red$ induces a ringed space
embedding $(X,\cA_\red)\rightarrow (X,\cA)$ whose underlying map of spaces is
the identity on points of $X$. The ringed space $(X,\cA_\red)$ is denoted by
$X_{\rm red}$ and called the {\em reduced} space associated with
$(X,\cA)$; this will also be identified with the corresponding (real or
complex) manifold. According to (1) in the definition, $X_{\rm red}$ is the ringed
space of smooth (resp. holomorphic) functions associated with a real (resp.
complex) manifold of real (resp. complex) dimension $m$. In the real case, we have $\cA_\red=\cC^\infty(X_\red)$ while in the
complex case we have $\cA_\red=\cO(X_\red)$. A supermanifold $X$ is called compact,
connected etc.\ if the underlying manifold $X_\red$ has the corresponding property.

Each of the local rings $\cA_x$ ($x\in X)$ is an augmented superalgebra, whose
augmentation morphism is the natural projection $\epsilon_x:\cA_x\rightarrow \cA_x/\gm_x=k$
($\gm_x$ is the unique maximal ideal of $\cA_x$ while $k=\R$ or $\C$ for real and complex supermanifolds, respectively). This is a
$k$-superalgebra morphism from $\cA_x$ to $k$, where the latter is viewed as a
commutative superalgebra over itself concentrated in degree zero. The
augmentation morphism is sometimes called the `body map', while
$\epsilon_x(f)$ is called the `body' of an element $f$ of $\cA_x$. When $X$
has dimension $(m|n)$, we have isomorphisms of superalgebras $\cA_x\cong k[\zeta^1\ldots \zeta^n]$ (the Grassmann $k$-algebra on $n$ odd
generators $\zeta^1\ldots \zeta^n$) for any $x\in X$. Furthermore, $\gm_x$ can
be identified with the maximal ideal $\langle \zeta^1\ldots \zeta^n\rangle$ of
this Grassmann superalgebra.

Condition (2) in the definition means
that $\hcA$ is the sheaf of smooth (resp. holomorphic) sections of the parity
change $\Pi E$ of a complex rank $n$ vector bundle $E$ over $X_\red$, which is
a holomorphic bundle in the complex supermanifold case. By convention, we will
denote $\cA$ by $\ucO$ respectively $\ucC$ for the case of complex resp. real
supermanifolds, and let $\cO$ respectively $\cC$ denote the corresponding reduced sheaves.

The following notations will be used later in this paper. For any point $x\in
X$, we let $\cA_x^\times$ denote the subgroup of invertible elements $\cA_x$,
i.e.\ those elements $f$ of $\cA_x$ such that $\epsilon_x(f)\neq 0$. We also
let $\cA_x^{\times, ev}$ denote the subgroup consisting of all even elements
of $\cA_x^\times$. Finally, we let $\cA^\times$ and $\cA^{\times,ev}$ be the
subsheaves of $\cA$ consisting of those elements whose stalk values at all
points $x$ belong to $\cA_x^\times$ and $\cA_x^{\times, ev}$ respectively. We
have sheaf inclusions $\cA^{\times, ev}\subset \cA^\times\subset \cA$. 
When $X$ is a real supermanifold, we let $\cC_{>0}$
denote the subsheaf of $\cC(X)$ consisting of `superfunctions with positive
body'. More precisely, we set $\cC_{>0}(U)=\{f\in \cC(U)|\epsilon_x(f(x))>0~~\forall x\in U\}$, where $U$ is any open subset of $X$ (here $f(x)\in \cC_x$ is
the stalk value of $f$ at $x$).  Notice that $\cC_{>0}$ is a subsheaf of $\cC^\times$.

\paragraph{Supervector bundles.}

Let $(X,\cA)$ be a supermanifold of dimension $(m|n)$ over $k=\R$ or $\C$ (i.e.\ a real or complex
supermanifold). A superfibration $E\stackrel{\pi}{\rightarrow} X$ is a fibration in the
category of supermanifolds over $k$, while a super fiber bundle is a fiber bundle in
that category. Such a fiber bundle is called a supervector bundle
of rank $(p|q)$ if its local trivializations over sufficiently small sets
are modeled on the bundle $U\times \A^{p|q}$ with $U\subset X$, while its transition functions are valued in the supergroup
${\rm GL}_k(p|q)$. Here $\A^{p|q}$ is the affine superspace of
dimension $(p|q)$ over $k$. The associated sheaf of sections is defined
through ${\cal E}=\Hom_{\cA}(\cA,\cA_{E,{\rm lin}})$, where $\cA_{E,{\rm lin}}$ is
the subsheaf of the structure supersheaf $\cA_E$ of $E$ whose local sections
are linear along the fibers of $E$. This sheaf is locally free of rank $(p|q)$
as a sheaf of $\cA$-supermodules. Conversely, any sheaf $\cE$ of $\cA$-supermodules
which is locally free of rank $(p|q)$ can be viewed as the sheaf of sections
of a supervector bundle of rank $(p|q)$ given by
$E=\Spec[S^\bullet_{\cA}(\cE^{\rm v})]$ (with the obvious projection) where
$\cE^{\rm v}=\Hom_{\cA}(\cE,\cA)$
is the dual sheaf and $S^\bullet_\cA $ is the functor on the category of sheaves
of $\cA$-supermodules induced by  taking the total graded symmetric algebra over a supermodule. 
Connections on super-vector bundles are defined by mimicking the classical
theory. 

\paragraph{Restriction and reduction of sheaves and supervector bundles.}

Given a sheaf $\cE$ of $\cA$-supermodules on $X$, its {\em restriction to $X_\red$} is the
sheaf on $X$ defined through:
\be 
\cE_{\rm res} := \cA\otimes_{\cA_\red} \cE\cong \cE/\cA_n\cdot \cE ~~. 
\ee 
This is a sheaf of $\cA_{\rm red}$-supermodules, i.e.\ sheaf of supermodules on
the ringed space $X_\red$. The subsheaf $\cE_\red:=(\cE)_{\rm res}^+$ of
even elements in $\cE_{\rm res}$ is called the {\em reduction} of $\cE$; it is
an ordinary sheaf of modules over $X_\red$.

When $\cE$ is locally free of rank $(p|q)$ with associated supervector bundle
$E\stackrel{\pi}{\rightarrow} X$, then $\cE$ is locally isomorphic with the free sheaf $\cA^{p|q}$ and
$\cE_{\rm res}$ is locally isomorphic with $\cA_{\rm
red}^{p|q}=\cA_\red^{\oplus p}\oplus (\Pi \cA_\red)^{\oplus q}$, thus locally
free as a sheaf of $\cA_{\rm red}$-supermodules and represented by a {\em
super}vector bundle $E_{\rm res}=E_{\rm res}^+\oplus E_{\rm res}^-\stackrel {\pi_{\rm res}}{\rightarrow} X_\red $ of rank
$(p,q)$ over $X_{\rm red}$, called the restriction of the supervector bundle
$E$. The image of a global section $s$ of $E$ though the projection
$\cE(X)\rightarrow \cE_{\rm res}(X)$ is denoted by $s_{\rm res}$ and called
the {\em restriction} of $s$. Notice that
$s_{\rm res}=1_{\cA(X)}\otimes_{{\cA}_\red(X)}s$.  The sheaf $\cE_{\rm red}$ is
again locally free and represented by the ordinary vector bundle
$E_\red=E_{\rm res}^+\rightarrow X_\red$ (the even subbundle of $E_{\rm  res}$) on $X_\red$. Notice that
the total space $E_\red$ is the underlying ordinary manifold of the total
supermanifold $E$ which is the total space of $E\stackrel{\pi}{\rightarrow} X$ . 
Also notice that when $E$ has rank $(p|0)$ (i.e.\ when $q=0$), then $E_\red=E_{\rm res}$. 

Notice that the sheaf of supersections of $E$ can be described as
$\cE=\cA\otimes_{\cA_\red} \cE_\res$. 

\paragraph{Natural sheaves and bundles.}

The {\em tangent sheaf} of $X$ is the sheaf $\cT_X:={\underline \Der}(\cA)$
of derivations of $\cA$. This is locally free of rank $(m|n)$. The super
vector bundle $T X$ associated with $\cT_X$ is called the {\em tangent bundle}
of $X$. Super-vector fields on $X$ are defined as global
supersections of $\cT_X$. The {\em cotangent sheaf} is the dual sheaf
$\cT_X^{\rm v}=\Hom_{\ucO}(\cT_X,\C)$, whose global supersections are the one-forms on
$X$. It is again locally free and represented by the {\em cotangent bundle} $T^*X$. 
Similarly, one defines the supertensor sheaves
$\cT_X\binom{p}{q}=\cT_X^{\otimes p}\otimes
(\cT_X^{\rm v})^{\otimes q}$, whose global sections are tensor superfields of type 
$\binom{p}{q}$ on $X$. These sheaves are
locally free and represented by the tensor bundles $\cT
\binom{p}{q}(X)$. The (locally
free) sheaf of $p$-forms is $\Omega^p_X:=\wedge^p {\cal T}^{\rm v}_X$,
where $\wedge$ is the graded wedge product; this sheaf is represented by the
$p$-form bundle $\Lambda^p T^* M$ (generally, one has $\Omega^p_X\neq 0$ for all $p\geq 0$).
The de Rham super differential $d$ is defined by mimicking the classical
construction. We also have the symmetric supertensor sheaves
$S^p_{\cA}(\cT_X)$, which are represented by the vector super bundle
$S^p( TX)$ etc. The reductions of all these sheaves and bundles are the
corresponding natural sheaves and bundles of $X_\red$. For example, we have
$(TX)_\red=T (X_\red)$ etc.

\subsection{Complex supermanifolds}

A complex supermanifold $(X,\ucO)$ of dimension $(m|n)$ has an underlying real
supermanifold $(X,\ucC)$ of dimension $(2m|2n)$ (see \cite{Haske-1987aa}). We
have a morphism of ringed spaces $(X,\ucC)\rightarrow (X,\ucO)$ whose underlying
map of spaces is the identity and whose sheaf map $\ucO\rightarrow \ucC$ is an
inclusion. There is a local isomorphism $\ucC\cong {\bar
\ucO}\otimes_{\cC} \ucO$ where the sheaf ${\bar \ucO}$ of antiholomorphic
superfunctions and the conjugation $\bar{~}:\ucO\rightarrow {\bar \ucO}$ are
defined \cite{Haske-1987aa} using the fact that $\ucO$
is locally the exterior algebra of an $\cO$-supermodule. Similarly, we have
a local isomorphism ${\hat \ucC}\cong {\hat {\bar \ucO}} \otimes_{\cC}
{\hat \ucO}$. As for ordinary complex manifolds, we have super-Dolbeault
decompositions: 
\beq \nn
\Omega^k_X=\oplus_{p+q=k}\Omega_X^{p,q}~~,  
\eeq
and the global sections of $\Omega_X^{p,q}$ are called $(p,q)$-forms on $X$.
Decomposing $d$ accordingly gives the Dolbeault super differentials $\partial$ and
${\bar \partial}$.

For any holomorphic supervector bundle $E$ of rank $(p|q)$ on $(X,\ucO)$, we let $\ucO(E)$ denote the sheaf of holomorphic supersections of $E$
and $\ucC(E)$ its sheaf of smooth supersections. These sheaves are locally free of
rank $(p|q)$ respectively $(2p|2q)$ over $\ucO$ and $\ucC$ respectively. We
have $\ucC(E)=\ucC\otimes_{\cC^\infty } \cC^\infty(E_\res)$ and 
$\ucO(E)=\ucO\otimes_{\cO} \cO(E_\res)$, where $\cC(E_\res)$
and $\cO(E_\res)$ are the ordinary sheaves of smooth resp. holomorphic
sections of the supervector bundle $E_\res$ over the ordinary manifold
$X_\red$. 

We let $H^0(E)=\ucO(E)(X)$
and $\Gamma(E)=\ucC(E)(X)$ denote the spaces of global holomorphic and smooth
supersections; these are supermodules over the superalgebras $\ucO(X)$ and $\ucC(X)$ respectively. When $X$
is compact, we have $\cO(X)=\C$ while $\cC(X)$ is infinite-dimensional as a
$\C$-vector space unless $X_\red$ consists of a finite set of points. In this case, $H^0(E)$ is a finite-dimensional vector
space while $\Gamma(E)$ is infinite-dimensional as a vector space unless
$X_\red$ consists of a finite set of points.

A {\em holomorphic super line bundle} on $(X,\ucO)$ is a holomorphic supervector bundle $L$ of
rank $(1|0)$. We say that $L$ is {\em positive} if $L_\red$ is positive
as an ordinary line bundle over $X$. Notice that any super line bundle satisfies
$L_\res=L_\red$. 

\paragraph{Hermitian structures.}

Let $E$ be a complex supervector bundle and $\cE=\cC(E)$ be its sheaf of
smooth supersections. A {\em super Hermitian pairing} $h$ on $E$ is a global supersection of
the sheaf $\Hom_{\ucC}({\overline \cE}\otimes_{\ucC}\cE,\ucC)$ such that its
value $h_p$ on the stalk at any point $p\in X$ is a super Hermitian pairing on the fiber
$E_p$. We say that $h$ is nondegenerate if each $h_p$ is. We say that $h$ is
positive-definite (or a Hermitian supermetric) if each $h_p$ is positive-definite, i.e.\ if $h_\red$ is a Hermitian {\em metric} on the bundle $E_\red$. A Hermitian supermetric $g$ on $T
X$ is called a Hermitian supermetric on $X$, in which case the pair $(X,g)$ is
called a Hermitian supermanifold (in this case, $(X_\red,g_\red)$ is a
Hermitian manifold).

\paragraph{K{\"a}hler supermanifolds.}

A {\em K{\"a}hler super form} on $X$ is a nondegenerate $(1,1)$-form $\omega$
such that $d\omega=0$ and such that $\omega_\red$ is positive-definite. By nondegeneracy, we mean that the stalk values $\omega_p$ are nondegenerate bilinear pairings on the vector superspaces $T_p X=\cT_{X,p}$ for each $p\in X$.  A
{\em K{\"a}hler supermanifold} is a Hermitian supermanifold $(X,g)$ such that
the 2-form $\omega_g:=i{\bar \partial} \partial g$ is a K{\"a}hler super form.

\paragraph{Projective superspaces.} Let $V=V_+\oplus V_-$ be a complex supervector
space with $\dim_\C V_+=m+1$ and $\dim_\C V_-=n$. The projectivisation of $V$ is the projective superspace $\P V=(\P
V_+, \wedge^\bullet [V_-\otimes O_{\P V_+}(-1)])$, viewed as a
(split\footnote{A supermanifold $(X,\cA)$ is called \emph{split} if the sheaf $\cA$ is globally isomorphic to
$\wedge^n_{\cA_\red}\hcA$ (rather than simply locally isomorphic).}) complex
supermanifold.  This comes endowed with a holomorphic super line bundle
$H:=O(1)$ called the hyperplane bundle, whose powers we denote by
$O(k):=O(1)^{\otimes k}$. The dual holomorphic superbundle
$O(-1)=O(1)^*$ is called the tautological super line bundle. A
$\Z_2$-homogeneous basis $e_0\ldots e_{m+n}$ of $V$ (with $e_0\ldots e_m\in
V_+$ and $e_{m+1}\ldots e_n\in V_-$) determines supercoordinates on the affine
supermanifold $\A_V$ associated with $V$, which in turn give a basis of the space of
global holomorphic supersections $z_0\ldots z_{m+n}$ of $O(1)$. The latter are
the {\em homogeneous supercoordinates} of $\P V$ determined by the given
homogeneous basis of $V$. The homogeneous supercoordinate ring
$\oplus_{k=0}^\infty{H^0(O(k))}$ (with multiplication given by the tensor
product and the $\Z_2$-grading induced from $O(k)$) is isomorphic as a
$\C$-superalgebra with the free supercommutative superalgebra $\C[z_0\ldots
z_{m+n}]$ generated by the homogeneous supercoordinates. This algebra also has
a $\Z$-grading given by the degree of monomials in $z_0\ldots z_{m+n}$,
and the supervector space $H^0(O(k))$ identifies with the component of
degree $k$ with respect to this grading.

A superscalar product $(~,~)$ on $V$ makes $\P V$ into a K{\"a}hler
supermanifold as follows. Since the total space of $O(-1)$ identifies with
the affine supermanifold $\A_V$ defined by $V$, the total space of $O(1)$ identifies
with the affine supermanifold $\A_{V^*}$ defined by $V^*$ and thus carries the Hermitian
metric induced by the superscalar product $(~,~)_*$. The former gives the Hermitian supermetric
$h$ on $O(1)$. This determines a K{\"a}hler supermetric on $\P V$, known as the
\emph{Fubini-Study supermetric} defined by $(~,~)$, through the relation: 
\beq
\omega=i\partial{\bar \partial} \ln h(z,z) ~~,  
\eeq
where the natural logarithm $\ln \alpha$  of an element of the Grassmann algebra $\C[z_{m+1},\ldots,$ $
z_{m+n}]$ is defined through its power series, when $\alpha$ is invertible in this superalgebra.

When $V=\C^{m+1|n}$ endowed with its canonical superscalar product, then the
projective superspace $\P V$ is denoted by $\P^{m|n}$ and called the
projective superspace of type $(m|n)$. This is isomorphic as a Hermitian
supermanifold to the projective superspace over any supervector space of
dimension $(m+1|n)$.

\subsection{Hodge supermanifolds and quantum super line bundles}

Consider a connected compact complex supermanifold $X$ of dimension
$(m|n)$. By definition, a {\em polarization} of $X$ is a positive holomorphic
super line bundle $L$ over $X$. The following {\em Kodaira superembedding
theorem} was proved in \cite{MR1032867}: Given a polarized K{\"a}hler supermanifold
$(X,L)$, there exists a positive integer $k_0$ such that the tensor powers
$L^k:=L^{\otimes k}$ are very ample for all $k\geq k_0$ in the sense that the
super Kodaira map defined by any homogeneous basis of the complex supervector
space $H^0(L^k)$ is a superembedding of $X$ in the projective superspace $\P
[H^0(L^k)^*]$.

As in the case of ordinary manifolds, there is the natural concept of a Hodge
supermanifold, providing a connection between K{\"a}hler and algebraic
supergeometry \cite{Kostant:1975qe}. A K{\"a}hler form $\omega$ is called {\em
integral}, if its cohomology class $[\omega]$ belongs to $H^2(X,\Z)$. In this
case, $(X,\omega)$ is called a {\em Hodge supermanifold}. It was shown by
Kostant (\cite{Kostant:1975qe}, Prop.\ 4.10.2) that in this case $X$ admits a
positive holomorphic super line bundle $L$ endowed with a
connection $\nabla$ such that 
$\omega=\frac{i}{2\pi} F_\nabla$ where $F_\nabla$ is the curvature of
$\nabla$ (in particular, we have $[\omega]=c_1(L)$). 
A triplet $(X,\omega,L)$ of this type is called a {\em polarized Hodge supermanifold}. 

Given a polarized Hodge supermanifold $(X,L,\omega)$, the super line bundle $L$
carries a Hermitian supermetric $h$ which determines the connection $\nabla$ as its
Chern connection, i.e.\ the unique connection of Dolbeault type $(1,0)$ compatible with $h$; in
fact, $h$ and $\nabla$ essentially determine each other
\cite{Tuynman:1992zm}. The quadruple $(X,L,h,\omega)$ is called a {\em
  prequantized Hodge supermanifold}. 
Under the replacement $L\rightarrow L^k$, we have an induced supermetric
$h_k=h^{\otimes k}$ on $L^k$ and the corresponding Chern
connection $\nabla_k=\nabla^{\otimes k}$ with curvature
$F_{\nabla_k}=\frac{k}{2\pi i}\omega$. Fixing a measure $\mu$ on $X$ yields
a superscalar product on the vector super space $H^0(L^k)$ \cite{Tuynman:1992zm}:
\begin{equation}\label{L2mu}
\langle s_1,s_2\rangle_k^{\mu,h}:=\int_{X}{d\mu ~h_k(s_1,s_2)} ~~.
\end{equation}

The standard choice for $d\mu$ is the super Liouville measure
$d\mu_\omega$ defined by the K{\"a}hler superform $\omega$. On a super-coordinate
chart $U$ with local coordinates $Z^I=z^1,\ldots,z^m,\zeta^1,\ldots,\zeta^n$, we have:
\begin{equation}
d\mu_\omega|_U:= (2\pi)^n|{\rm sdet}(\omega_{IJ})|dz^1\wedge
 d\bz^1\wedge\ldots\wedge dz^m\wedge d\bz^m i d\zeta^1d \bar{\zeta}^1\ldots i d\zeta^nd
 \bar{\zeta}^n~,
\end{equation}
where $\omega_{IJ}$ are the coefficients of the K{\"a}hler form
$\omega=\omega_{IJ}dZ^I\wedge d\bar{Z}^J$ and ${\rm sdet(A)}$ is the
superdeterminant (Berezinian) of the supermatrix $A$. However, it is
is desirable to work in a more general setting in order to include e.g.\ the case when $X$ is
algebraically a Calabi-Yau supermanifold and $\mu_{CY}$ is the volume form
determined by its the holomorphic volume element.

\paragraph{Remark.} As for ordinary manifolds, we can derive a
local K{\"a}hler potential $K$ from the Hermitian bundle supermetric $h$. Given a 
global supersection $\sigma$ of $L$, we let  $K_\sigma:=-\log h(\sigma,\sigma)$, which
is a K{\"a}hler potential on the set $U_{\sigma}:=\{x\in X|\sigma(x)\in
{\ucO}_x^\times\}$ (here $\ucO_x^\times$ is the supercommutative group of
invertible elements in the local superalgebra $\ucO_x$): 
\begin{equation}
\omega=\frac{i}{2\pi}\partial {\bar \partial} K_{\sigma}=\frac{1}{2\pi i} \log
h(\sigma,\sigma)=\frac{i}{2\pi} F_\nabla~.
\end{equation}

\subsection{Parameterizing Hermitian bundle supermetrics and polarized K{\"a}hler super forms}

Let us fix a polarized complex supermanifold $(X,L)$, where
$L\stackrel{\pi}{\rightarrow} X$ is a positive holomorphic super line bundle on
$X$. Since $L$ has rank $(1,0)$, we have that $L_{\rm res}=L_{\rm
  red}\stackrel{\pi_\red} \rightarrow X_\red$ is an
ordinary holomorphic line bundle on the complex manifold $X_\red$.
We let $L_\red^\times\stackrel{\pi^\times_\red}{\rightarrow} X_\red $ be the total
space $L_\red$ with the zero supersection removed. A Hermitian supermetric $h$ on $L$ is determined by the associated
Hermitian $\C$-sesquilinear maps $h_x:(L_\red)_x\times (L_\red)_x\rightarrow
\ucC_x$ on the fibers of $L_\red$ at the points $x$ of $X_\red$.
Hence $h$ is uniquely determined by the global supersection ${\hat  h}$ of
$(\pi_\red^\times)^*(\ucC)$ given by: 
\beq
\label{eqn:hsquarenorm}
{\hat h}(q):=h(q,q)\in \cC_{\pi^\times_\red(q)}~~,~~q\in L_\red^\times~~. 
\eeq
where $\ucC$ is the sheaf of smooth superfunctions
on $X$, viewed as a sheaf of $\cC$-superalgebras on $X_\red$. Notice that the right hand side is invertible in the superalgebra
$\ucC_{\pi_\red(q)}$. We have ${\hat h}(cq)=|c|^2{\hat
  h}(q)$ for all $q\in L_\red^\times$ and all $c\in \C^\ast$. The set  ${\rm Met}(L)$ of
Hermitian supermetrics on $L$ can be identified with the set of all such
${\hat h}$. If we fix a reference super metric $h_0$ on $L$, then any other
supermetric $h$ is described by the global supersection $\phi=\frac{{\hat
    h}}{{\hat h}_0}$ of $\ucC$, which is a smooth superfunction on $X$ whose
body (projection to $\cC(X)$) is a positive-definite ordinary smooth
function. We thus find that ${\rm
  Met}(L)$ can be identified with the real cone $\ucC_{>0}(X)$ of all such
smooth superfunctions. 

In this paper, we will use a slightly different parameterization 
when $L$ is very ample. For any $q\in L_\red^\times$, we let ${\hat
q}:H^0(L)\rightarrow \ucO_{\pi(q)}$ be the $\C$-linear functional (called {\em
evaluation functional}) defined through: 
\beq
\label{hatq} s(\pi(q))={\hat q}(s)q~,~~~s\in H^0(L)~. 
\eeq 
We have the obvious property
$\widehat{cq}=\frac{1}{c}{\hat q}$ for all non-vanishing complex numbers
$c$. The very ampleness of $L$ implies ${\hat q}\neq 0$ for all $q\in
L_\red^\times$.

Consider a Hermitian superscalar product $(~,~)$ on $E:=H^0(L)$, where
$\dim_\C(E)=(m+1|n)$. Picking a homogeneous basis
$s_0,...,s_{m+n}$ of $E$ with $s_0\ldots s_m$ even and $s_{m+1}\ldots s_{m+n}$ odd, we
let $G_{ij}=(s_i,s_j)\in \C$.  Also let $G^{ij}$ be the
inverse of $G_{ij}$, so that $\sum_{j=0}^{m+n}G^{ij}G_{jk}=\delta^i_k$. Since
the superscalar product is even, the matrix $G$ is block-diagonal: 
\beq 
\nn 
G_{i\iota}=G_{\iota i}=0~~{\rm for}~~i=0\ldots m,~~\iota=m+1\ldots m+n~~.  
\eeq
The super-Hermitian property of $(~,~)$ reads: 
\bea
&&{\overline G_{ij}}=G_{ji}~~~~~i,j=0\ldots m\nn\\
&&{\overline G_{\iota \gamma}}=-G_{\gamma \iota}~~~\iota,\gamma=m+1\ldots m+n\nn~~.
\eea 
Furthermore, the submatrix $(G_{ij})_{i,j=0\ldots m}$ is positive-definite.

Let us fix a point $q\in L_\red^\times$ with $\pi(q)=x$.
As explained in Section 2.1., the pairing $(~,~)$ induces a superscalar product $(~,~)_*$ on the dual space
$H^0(L)^*=\Hom_\C(H^0(L),\C)$. The latter extends uniquely to a Hermitian bilinear form $\lbr~,\rbr_x$ on
the $\ucO_{x}$-module $\Hom_\C(H^0(L),\ucO_x)=H^0(L)^*\otimes_\C
\ucO_x$. This allows us to consider the
Hermitian supermetric $h_B$ on $L$ whose `square norm' superfunction is given
by (this is well-defined since the denominator is invertible in the local
algebra $\cO_{x}$):
\beq
\label{hatbergman} {\hat h}_B(q)=\lbr {\hat q},{\hat
q}\rbr_{x}=\frac{1}{\sum_{i,j=0}^{m+n}G^{ij}\overline{\hat{q}(s_i)}\hat{q}(s_j)}\in \ucO_{x}~~~~~,~~~~ q\in L_\red^\times,\ x=\pi(q)~~. 
\eeq
This is called the {\em Bergman supermetric}
on $L$ defined by $(~,~)$. Since we now have a
reference Hermitian supermetric on $L$, we can describe any other supermetric $h$
via the superfunction: 
\beq
\label{eqn:relepsilon} \epsilon:=\frac{{\hat h}}{{\hat h}_B}\in \ucC_{>0}(X)~~, 
\eeq
which we call the {\em epsilon superfunction of $h$ relative to $(~,~)$}: \beq
\label{he} h(q,q)=\epsilon(\pi(q))h_B(q,q)~~.  \eeq More explicitly, we have
$\epsilon(x)=\sum_{i,j=0}^{m+n} {G^{ij}h(x)(s_i(x),s_j(x))}$. Thus, Hermitian
supermetrics on $L$ are parameterized by their relative epsilon superfunctions,
once one fixes a superscalar product on $H^0(L)$.

The relative epsilon superfunction defined above depends on $h$ and on the superscalar
product chosen on $H^0(L)$ and is a generalization of the more familiar
object considered in
\cite{Rawnsley:1976gb,Rawnsley:1990tj,Rawnsley-1993aa}. To make contact with
the latter, notice that fixing $h$ gives a distinguished choice of a superscalar 
product on $H^0(L)$, namely the $L^2$ superscalar product $\langle~,~\rangle$
defined by $h$ and by the Liouville density of the associated super K{\"a}hler
form $\omega$. The epsilon superfunction of $h$ with respect to this
superscalar product depends only on $h$ (remember that $\omega$ is
determined by $h$), and will be called the {\em absolute epsilon
superfunction} of $h$. The latter generalizes the absolute epsilon function
considered in \cite{Rawnsley:1976gb, Rawnsley:1990tj, Rawnsley-1993aa}.

The $L$-polarized K{\"a}hler supermetric on $X$ determined by $h_B$ is called
the {\em Bergman supermetric on $X$ induced by $(~,~)$}. Its K{\"a}hler super form
is denoted by $\omega_B$. The K{\"a}hler super form $\omega$ determined by the
Hermitian bundle supermetric (\ref{he}) takes the form: \beq \nn
\omega=\omega_B-\frac{i}{2\pi}\partial{\bar \partial}\log \epsilon~~, \eeq so
as expected we have $\omega=\omega_B$ iff the relative epsilon superfunction of $h$
is constant. Since $\omega$ determines $h$ up to multiplication by a constant,
it also determines the relative epsilon superfunction of the latter up to the same
ambiguity.  We shall see below that $L$-polarized Bergman supermetrics are those
supermetrics induced on $X$ by pulling-back Fubini-Study supermetrics through the
Kodaira superembedding $i:X\hookrightarrow \P^{m|n}$ (where
$\dim_\C(H^0(L))=(m+1|n)$) determined by the very ample super line bundle $L$,
where the Fubini-Study supermetric being pulled-back is determined by the superscalar
product on $H^0(L)^*$.

\paragraph{Remark.} The Hermitian superscalar product on $H^0(L)$ defined by
$h_B$ and by the volume form of $\omega_B$: \beq \nn \langle
s,t\rangle=\int_{X} d\mu_{\omega_B} h_B(s,t)~~(s,t\in H^0(L)) \eeq {\em need
not} coincide with the superscalar product $(~,~)$ which parameterizes $h_B$. If
they do, then one says that the superscalar product $(~,~)$ and associated Bergman
bundle and manifold supermetrics $h_B$, $\omega_B$ are {\em balanced}
\cite{Donaldson:2001aa,Catenacci:arXiv0707.4246}. Clearly, for $\omega_B$ to
be balanced, the {\em absolute} epsilon superfunction has to be constant.

\subsection{Bergman supermetrics from metrized Kodaira superembeddings}

Let $X$ be a compact complex supermanifold. By the Kodaira superembedding
theorem, a positive super line bundle $L$ on $X$ gives a holomorphic
superembedding $i:X\hookrightarrow \P V$, where $E:=H^0(L)$ and $V=E^*$ is the
space of holomorphic supersections of $L$, whose complex superdimension we denote
by $(m+1|n)$.  The embedding allows us to view $X$ as a nonsingular projective
supervariety in $\P V$, whose homogeneous coordinate ring $R(X,L)=\oplus_{k\geq
0}{H^0(L^k)}$ is generated in monomial degree $k=1$. In particular, $L$ and the pull-back
$i^*(H)$ of the hyperplane superbundle $H:=O_{\P V}(1)$ are isomorphic as
holomorphic super line bundles.

Conversely, if we are given any smooth projective supervariety $X$ in a projective
superspace $\P V $ whose vanishing ideal $I(X)$ is generated in monomial degrees greater
than one, then the restriction $O_X(1)=O_{\P V}(1)|_X$ is very
ample and the embedding $X\embd \P V$ can be viewed as the Kodaira superembedding
determined by this restriction. The space of holomorphic supersections of $O_X(1)$ identifies with the supervector space $E=V^*$.

 A {\em metrized Kodaira embedding} is a Kodaira superembedding
determined by a very ample super line bundle $L$ on $X$ together a fixed choice of a
Hermitian superscalar product $(~,~)$ on its space of holomorphic supersections
$E:=H^0(L)$. For such embeddings, the superscalar product on $E$ induces a superscalar
product on $V=E^*$, which makes $\P V $ into a (finite-dimensional) projective
super Hermitian space. The later carries the Fubini-Study
supermetric\footnote{Homogeneous 
K{\"a}hler supermetrics on $\P V $ are in bijection with super Hermitian
scalar products on $E$ taken up to constant rescaling, and these are the
Fubini-Study supermetrics. They are all related by ${\rm PGL}(E)$-transformations.}
determined by the superscalar product.  Its K{\"a}hler super form is given by: \beq \nn
\pi^*(\omega_{FS})(v)=\frac{i}{2\pi}\partial {\bar \partial}\log[(v,v)_*]~~,
\eeq where $\pi:V \rightarrow \P V$ is the canonical projection while
$(~,~)_*$ is the superscalar product induced on $V=E^*$.  There exists a one to one correspondence
between metrized Kodaira superembeddings of $X$ and holomorphic superembeddings in
finite-dimensional projective super Hermitian spaces such that the vanishing ideal of
the superembedding is generated in monomial degrees greater than one.

The Fubini-Study supermetric admits the hyperplane superbundle $H$ as a quantum line
superbundle, when the latter is endowed with the Hermitian bundle supermetric $h_{FS}$
induced from $E$. Since $L\simeq i^*(H)$ as holomorphic super line bundles, the
pull-back $i^*(h_{FS})$ defines a Hermitian supermetric $h_B$ on $L$. The latter
coincides with the Bergman bundle supermetric determined by $(~,~)$. The
pulled-back K{\"a}hler form $\omega_B=i^*(\omega_{FS})$ admits $(L,h_B)$ as a
quantum super line bundle, and coincides with the Bergman K{\"a}hler form determined
by $(~,~)$. It follows that Bergman supermetrics on $X$ coincide with pull-backs of
Fubini-Study supermetrics via metrized Kodaira embeddings.

\paragraph{Remark.} A choice of homogeneous basis $z_0\ldots z_{m+n}$ for $E=V^*$ allows us to
express $v\in V$ as: $v=\sum_{i=0}^{m+n}{v_ie_i}$, where $(e_i)$ is the
homogeneous basis of
$V$ dual to $(z_i)$ and $v_i=z_i(v)$. This gives an identification of $V$ with
the supervector space $\C^{m+1|n}$ endowed with the superscalar product given by $\langle
u,v\rangle=\sum_{i,j=0}^{m+n} G^{ij}{\bar u}_i v_j$, where the $G^{ij}$ are given
as above.  Then $\P V$ identifies with $\P^{m+1|n}$ endowed with the Fubini-Study
supermetric defined by this superscalar product. It is customary to choose an
orthonormal basis, in which case the Fubini-Study supermetric takes the familiar
form in homogeneous supercoordinates. In this case, the freedom of choosing the
superscalar product $(~,~)$ is replaced by the freedom of acting with ${\rm PGL}_\C(m+1|n)$
transformations on the homogeneous supercoordinates of $\P^{m|n}$.  

\section{Generalized Berezin and Toeplitz quantization of supermanifolds}

In this section, we extend the generalized Berezin and Toeplitz quantization
procedure of \cite{IuliuLazaroiu:2008pk} to the case of K{\"a}hler
supermanifolds. The subtle point of this extension is to control the various
super Hermitian forms involved in defining the Rawnsley supercoherent projectors.

In the following, let $X$ be a K{\"a}hler supermanifold endowed with a fixed
very ample super line bundle $L$. We also fix a Hermitian superscalar product on
the finite-dimensional supervector space $E=H^0(L)$, whose dimension we denote
by $(m+1|n)$.

We will consider a homogeneous basis $s_0\ldots s_{m},s_{m+1}\ldots s_{m+n}$
of $E$, where $s_0\ldots s_m$ are even and $s_{m+1}\ldots s_{m+n}$ are odd. We
let $G$ be the Hermitian matrix with entries $G_{ij}:=( s_i, s_j)$
and $G^{ij}$ be the entries of the inverse matrix $G^{-1}$. Any supersection $s\in
E$ can be expanded as: \beq \nn s=\sum_{i,j=0}^{m+n}{G^{ij}( s_j,s) s_i}~~.
\eeq 
As mentioned above, the matrix $G$ is block-diagonal because the bilinear form
$(~,~)$ is even, and the submatrix $(G_{ij})_{i,j=0\ldots m}$ is
positive-definite. 

\subsection{Supercoherent states}

Recall that $\ucO(L)$ denotes the sheaf of holomorphic supersections of $L$. 
For any point $x\in X$, the stalk $\ucO_x(L)$ of this sheaf at $x$ is a free $\ucO_x$-supermodule
of rank $(1|0)$. We let $\ucO_x^{\times,ev}(L)$ be the set of even bases of this module, i.e.\
the set of even elements $q\in L_x$ such that $\ucO_x(L)=\ucO_x q$. We have
$\ucO_x^{\times, ev}(L)=\ucO_x^{\times,ev} q$ for any $q\in (L_\red^\times)_x$, where $\ucO_x^{\times,ev}$ is
the set of {\em even} invertible elements of $\ucO_x$ (this is a subgroup of
the multiplicative monoid of the superalgebra $\ucO_x$).

Given $q\in (\L_\red^\times)_x$ and a supersection $s\in E=H^0(L)$, we have: \beq
s(x)={\hat q}(s)q \eeq for some element ${\hat q}(s)\in \ucO_x$. This gives an
even $\C$-linear functional ${\hat q}:E\rightarrow \ucO_x$. Consider the
$\ucO_x$-supermodule $E_x:=\ucO_x\otimes_\C E$. The superscalar product on $E$
extends to a nondegenerate and even $\ucO_x$-sesquilinear map $\lbr~,~\rbr_x:E_x\times
E_x\rightarrow \ucC_x$ as follows: 
\beq \label{ext}
\lbr\alpha\otimes s,\beta \otimes
t\rbr_x= (-1)^{\tilde{s}\tilde{\beta}}({\bar \alpha}\beta)\otimes_\C (s,t)~~.  \eeq where $\alpha,\beta\in
\ucO_x$ and $s,t\in E$. These extended even pairings make each $E_x$ into a
Hermitian $\ucO_x$-module.

By the Riesz theorem, we have a uniquely determined
element $e_q\in E_x$ such that: \beq \lbr e_q,s\rbr_x={\hat q}(s)~~\forall
s\in E~~,  \eeq where we consider $s\in E$ tensored with the identity $1_{\ucO_x}$ in $E_x$. Direct computation gives the explicit expression: \beq \nn
e_q=\sum_{i,j=0}^{m+n}{G^{ji}\overline{{\hat q}(s_i)} \otimes_\C s_j}~~, \eeq
which implies: \beq \nn \lbr
e_q,e_q\rbr_x=\sum_{i,j=0}^{m+n}{G^{ij}\overline{{\hat q}(s_i)} {\hat
q}(s_j)}\in \ucO_x~~.  \eeq Notice that $e_q$ cannot be the zero supersection, since
that would imply that all supersections of $L$ (and thus of $L_\red$) vanish at
$x$, which is impossible since $L_\red$ is very ample. Also notice that $\lbr
e_q,e_q\rbr_x$ belongs to $(\ucC_{>0})_x\subset \ucC_x^\times$. Indeed, the fact that ${\hat q}$ is
an even map implies that $e_q$ itself is even and we have: \beq \ev_x(\lbr
e_q,e_q\rbr_x)=\sum_{i,j=0}^{m}{G^{ij}\overline{\ev_x({\hat q}(s_i))}
\ev_x({\hat q}(s_j))}~~, \eeq which is positive since the restriction of
$(~,~)$ to $E_+$ is positive-definite and because $L_\red$ is very ample. The
element $e_q$ of $E_x$ will be called the {\em Rawnsley supercoherent vector} defined
by $q$. This generalizes the coherent vectors introduced in
\cite{Rawnsley:1976gb} to the supermanifold case.

If $q'$ is another element of $(L_\red^\times)_x$, then $q'=cq$ for some $c\in
\C$ and we have $e_{q'}=\frac{1}{\bar c}e_q$. It follows that the
rank $(1|0)$ $\ucO_x$-module $l_x:=\langle e_q\rangle=\ucO_x e_q\subset E_x$
depends only on the point $x\in X$.  This can be interpreted as follows. Let
$\bar{L}$ be the super line bundle obtained by reversing the complex structure
of all fibers; this is a holomorphic super line bundle over the complex
supermanifold $\bar{X}$ obtained by reversing the complex structure of
$X$. The scaling property of supercoherent vectors implies that the element
$e_x:={\bar q}\otimes_\C e_q\in \bar{L}_x\otimes_\C E$ depends only on the point
$x\in X$.  The superscalar product on $E$ extends to a sesquilinear map $\lbr~,~\rbr$ taking
$[\bar{L}_x\otimes_\C E]\times [{\bar L}_y\otimes_\C E]$ into ${\bar
L}_x\otimes_\C {\bar L}_y$. So in particular, the combination $K(x,y)=\lbr e_x,e_y\rbr$
defines a holomorphic supersection $K$ of the external tensor product
$\bar{L}\boxtimes \bar{L}$ (which is a holomorphic super line bundle over
the supermanifold ${\bar X}\times {\bar X}$). This will be called the {\em reproducing kernel} of the
finite-dimensional super Hermitian space $(E, (~,~))$.

Rawnsley's {\em supercoherent projectors} are the $\ucO_x$-linear 'orthoprojectors' $P_x\in$\linebreak $
\End_{\ucO_x}(E_x)$ on the rank one submodules $l_x\subset E_x$: 
\beq\label{Rw}
P_x({\bf s})=\frac{e_q \lbr e_q, s\rbr_x}{\lbr e_q,e_q\rbr_x}\in l_x
~~~(s\in E_x)
\eeq
These are well-defined since $\lbr e_q,e_q\rbr_x$ is an
even invertible element of $\ucC_x$.  The projectors depend only on $L$, on the
point $x\in X$ and on the superscalar product chosen on $E$. Given a $\C$-linear
operator $C\in \End(E)$, its {\em lower Berezin symbol} is the smooth
superfunction $\sigma(C)\in \ucC(X)$ given by: \beq
\label{symbol} \sigma(C)(x):=\str(CP_x)=\frac{\lbr e_q,C e_q\rbr_x}{\lbr
e_q,e_q\rbr_x}~~.  \eeq This gives a $\C$-linear map $\sigma:\End(E)\rightarrow
\ucC(X)$, whose image we denote by $\Sigma$. Notice that $\sigma$ and
$\Sigma$ depend only on $L$ and on the superscalar product $(~,~)$ chosen on
$E$. The obvious property: \beq \nn \sigma(C^\dagger)=\overline{\sigma(C)}~~
\eeq implies that $\Sigma$ is closed under the complex conjugation of the
superalgebra $\ucC(X)$, i.e.\ we have 
$\bar{\Sigma}=\Sigma$.  Also notice that $\Sigma$ contains the constant unit
function $1_X=\sigma(\id_E)$ and that $\sigma$ is an even map: \beq
\widetilde{\sigma(C)}=\tilde{C}~~, \eeq when $C$ is a $\Z_2$-homogeneous
operator in $E$. It follows that $\Sigma$ is a $\Z_2$-homogeneous subspace of
$\ucC(X)$, i.e.\ $\Sigma=\Sigma_+\oplus \Sigma_-$ with $\Sigma_\pm=\Sigma\cap
\ucC(X)_\pm$.

\subsection{Generalized Berezin quantization}

As for ordinary manifolds, the Berezin symbol map $\sigma:\End(E)\rightarrow
\ucC(X)$ is injective so its kernel is trivial. This is
easily seen from expanding
\begin{equation} \lbr e_q, C
e_q\rbr_x=\sum_{i,j,k,l}(G^{ji}\hat{q}(s_i))^*G^{kl}\hat{q}(s_l)(s_j,C s_k)~.
\end{equation} As $P_x$ is independent of the choice of $q$ for every $x$, we
choose $q$ to be $(1,x)$ everywhere. Also note that we can write $C$ in the
above equation as $C=\sum_{i,j} |s_i)C^{ij}(s_j|$. Altogether we thus obtain
\begin{equation} \lbr e_q, C e_q\rbr_x=\sum_{i,j}C^{ij}\bar{s}_i(x)s_j(x)~. \label{eqn:opproduct}
\end{equation} 
As the $(s_i)$ form a (holomorphic) basis of $E$, it follows that if \eqref{eqn:opproduct} equals zero then $C^{ij}=0$.
 
It follows that the corestriction $\sigma|^\Sigma:\End(E)\rightarrow \Sigma$ is an
isomorphism of supervector spaces and we can associate an operator on $E$ to every
superfunction $f\in \Sigma$ via the {\em Berezin quantization map}
$Q=(\sigma|^\Sigma)^{-1}:\Sigma\rightarrow \End(E)$: \beq
\label{eqn:gen_berezin_q} Q(f):=\sigma^{-1}(f)~~~~\forall f\in \Sigma~~.  \eeq
The quantization map $Q$ is even and depends only on $L$ and on the choice of
superscalar product on $H^0(L)$.  It satisfies the relations: \beq \nn
Q(\bar{f})=Q(f)^\dagger~~,~~Q(1_X)=\id_E~~.  \eeq

\paragraph{The Berezin superalgebra.}  The {\em Berezin product}
$\diamond:\Sigma\times \Sigma\rightarrow \Sigma$ is defined via the formula:
\beq
\label{Ber_product} f\diamond g:=\sigma(Q(f)Q(g))\Leftrightarrow Q(f\diamond
g)=Q(f)Q(g)~~.  \eeq Together with the complex conjugation of smooth
superfunctions $f\rightarrow \bar{f}$, it makes $\Sigma$ into a unital
finite-dimensional associative $*$-superalgebra (in particular, the
conjugation of $\ucC(X)$ restricts to an even involution of $\Sigma$). The
Berezin quantization map gives an isomorphism of $*$-superalgebras: \beq \nn
Q:(\Sigma,\diamond,\bar{~})\rightarrow (\End(E),\circ,\dagger)~~.  \eeq Recall
that $(\End(E),\circ,\dagger )$ is a $*$-superalgebra with nondegenerate trace
given by the usual supertrace. It follows that the induced linear map (called
the {\em Berezin supertrace}): \beq
\label{strace} \vint{f}:=\str~Q(f)~~(f\in \Sigma) \eeq is a nondegenerate
supertrace on the Berezin superalgebra $(\Sigma,\diamond,\bar{~})$:
\begin{eqnarray} \vint{\bar f}&=&\overline{\vint f} ~~,\nn\\ \vint f\diamond g
&=&(-1)^{\tilde{f}\tilde{g}}\vint g\diamond f~~,\nn\\ \vint f\diamond g =0,\
&\forall& g\in \Sigma \Rightarrow f=0~~.\nn
\end{eqnarray}

The super Hermitian pairing on $\Sigma$ (called the {\em Berezin pairing})
obtained by transporting the super Hilbert-Schmidt pairing: \beq
\label{diamond_product} \prec f, g\succ_B:=\langle Q(f),
Q(g)\rangle_{HS}=\str\left(Q(f)^\dagger Q(g)\right) \eeq coincides with the
pairing induced by the Berezin supertrace: \beq \nn \prec f, g\succ_B=\vint
{\bar f}\diamond g~~.  \eeq Notice that $\prec 1_X, 1_X\succ_B=\langle
\id_E,\id_E\rangle_{HS}=m-n+1$, where $(m+1|n)$ was the superdimension of $E$.

\paragraph{The squared two point superfunction.} For later reference, we
define the superanalogue of the squared two-point function of coherent
states. For this, note that the $\ucO_x$-sesquilinear maps $\lbr~,~\rbr_x$ on $E_x$ (see equation \eqref{ext}), uniquely extend further to a pairing $\lbr~,~\rbr_{x,y}:[\ucO_x\otimes_\FC E]\times
[\ucO_y\otimes_\FC E]\rightarrow \bar{\CO}_x\otimes\CO_y$. Define the two-point superfunction $\Psi:X\times X\mapsto \ucC_x\otimes\ucC_y$ via:
\beq
\label{two_point}
\Psi(x,y):=\str(P_xP_y)=\sigma(P_x)(y)=\sigma(P_y)(x)=\frac{\lbr
  e_x,e_y\rbr_{x,y}\overline{\lbr e_x,e_y\rbr}_{x,y}}{\lbr e_x,e_x \rbr_x \lbr e_y,e_y\rbr_y}~~.
\eeq
As the supercoherent state projectors $P_x,P_y$ are even operators on $E$, $\Psi$ is symmetric on $X\times X$:
\beq
\nn
\Psi(x,y)=\Psi(y,x)~~\forall x,y\in X~~
\eeq
and vanishes at points $(x,y)$ where the directions of the supercoherent vectors
  $e_x$ and $e_y$ in $E$ are orthogonal to each other with respect to the
  pairing $\lbr~,~\rbr_{x,y}$.

\subsection{Changing the superscalar product in generalized Berezin quantization}

Any Hermitian superscalar product $(~,~)'$ on $E$ has the form: 
\beq
\label{productprime}
( s,t)'=( As,t )
\eeq
with $A$ a $(~,~)$-super Hermitian even invertible operator. Such an operator
has the block diagonal structure $A=A_+\oplus A_-$ (with
$A_\pm\in GL(E_\pm)$) where $A_\pm$ are $(~,~)_\pm$-Hermitian and $A_+$ is
positive definite. The supercoherent states with respect to the new 
product $(~,~)'$, which in turn induces the pairing $\lbr~,~\rbr_x'$, are given by:
\beq
\label{eprime}
e_q'=A^{-1}e_q~~(q\in L_x^\times)~~,
\eeq
while the new supercoherent projectors take the form: 
\beq
\label{Pprime}
P'_x=\frac{1}{\sigma(A^{-1})(x)}A^{-1}P_x~~(x\in X)~~.
\eeq
The symbol $\sigma(A^{-1})(x)=\frac{\lbr e_q|A^{-1}|e_q\rbr_x}{\lbr
  e_q|e_q\rbr_x}$ of $A^{-1}$ computed with respect to
$\lbr ~,~\rbr_x$ and the symbol $\sigma'(A)(x)=\frac{\lbr
  e'_q|A|e'_q\rbr_x'}{\lbr e'_q|e'_q\rbr_x'}$ of $A$ computed with
respect to $\lbr ~,~\rbr_x'$ are related by:
\beq
\label{rel1}
\sigma(A^{-1})(x)=\frac{1}{\sigma'(A)(x)}~~.
\eeq
As $A_+$ is positive definite, the body of both $\sigma(A)$ and $\sigma'(A)$
is non-vanishing. 
Therefore, both these superfunctions are invertible on $X$. Given an operator $C$, we have more generally:
\beq
\label{sigma_change}
\sigma'(C)=\frac{\sigma(CA^{-1})}{\sigma(A^{-1})}
\eeq
and:
\beq
\label{sigma_change2}
\sigma(C)=\frac{\sigma'(CA)}{\sigma'(A)}~~.
\eeq
Since $A$ is even, so are $\sigma'(A)$ and $\sigma(A^{-1})$. Therefore, the order
of multiplication in the fractions is irrelevant. Let $Q'$ be the Berezin
quantization map defined by $(~,~)'$ and $\Sigma'\subset \ucC(X)$
be the image of $\sigma'$. Equation (\ref{sigma_change}) shows that
\beq
\nn
\Sigma'=\frac{1}{\sigma(A^{-1})}\cdot \Sigma=\left\{\frac{1}{\sigma(A^{-1})}\cdot f~|~f\in \Sigma \right\}
\eeq
and that:
\beq
\nn
Q'(f)=Q(\sigma(A^{-1})f) A~~~\forall f\in \Sigma'~~.
\eeq

As for ordinary manifolds, we have the following proposition, whose proof
mimics that of the corresponding result of \cite{IuliuLazaroiu:2008pk}:

\paragraph{Proposition.} The Berezin quantizations defined by two different
superscalar products on $E$ agree iff $A$ is proportional to the
identity, i.e.\ iff the two superscalar products are related by a constant
scale factor $\lambda\in\C^*$. In this case, the supercoherent states differ
by a constant homothety and the Rawnsley supercoherent projectors are equal.

\subsection{Integral representations of the superscalar product}

In the following, let $L$ be a very ample super line bundle, $E=H^0(L)$ the
vector space of supersections and $(~,~)$ a superscalar product on $E$. We
will look for those superscalar products $(~,~)$ which admit
integral representations through a measure $\mu$ and a Hermitian supermetric
$h$ on $L$; such a representation is required for defining generalized
Toeplitz quantization.

A Hermitian bundle supermetric $h$ on $L$ can be
parameterized by its epsilon superfunction relative to $(~,~)$: \beq
\label{epsilon}
\epsilon(x):=h(x)(q,q)~\lbr
e_q,e_q\rbr_x=\sum_{i,j=0}^{m+n}{G^{ij}h(x)(s_i(x),s_j(x))}~~.  \eeq Note that
the right hand side is indeed independent of the choice of $q$. Furthermore,
$h(x)$ is uniquely determined by $h(q,q)$; conversely, the epsilon
superfunction determines a Hermitian supermetric via
$h(x)(q,q)=\frac{\epsilon(x)}{\lbr e_q,e_q\rbr_x}$.

Let us look for integral representations of $(s,t)$ 
of the following form: 
\beq 
\nn 
( s, t)=\int_Xd\mu(x)~h(x)(s(x),t(x))~~.  
\eeq
Since the right hand side equals $\int_X{d\mu(x) \epsilon(x)\lbr s, P_x t \rbr_x}$
we have:

\paragraph{Proposition.} The superscalar product $(~,~)$ on $E$ coincides
with the $L^2$ product induced by $(\mu,h)$ iff the relative epsilon
superfunction of the pair $(h,(~,~))$ satisfies the identity
\beq\label{completeness} \int_{X}{d\mu(x) {\epsilon}(x) P_x}=\id_E~~, \eeq
i.e.\ iff the supercoherent states defined by $(~,~)$ form an `overcomplete set'
with respect to the measure $\mu_{\epsilon}=\mu\epsilon $.

\

The precise mathematical meaning of equation (\ref{completeness}) is as
follows. Recall that the supercoherent projectors $P_x$ are $\ucO_x$-linear operators
acting in the free modules $E_x=\ucO_x\otimes_\C E$, i.e.\ elements of the free
$\ucO_x$-module $\End_{\ucO_x}(E_x)\cong \ucO_x\otimes_\C \End(E)$. 
The map $P$ which associates the operator $P_x$ to
every point of $x$ ($P(x):=P_x$) is a holomorphic supersection of the trivial bundle $X\times
\End(E)$, i.e.\ an element of the free $\ucO(X)$-module $\ucO(X)\otimes_\C
\End(E)$, while its product with the epsilon superfunction is a smooth
supersection of the same bundle and thus an element of the free
$\ucC(X)$-module  $\ucC(X)\otimes_\C
\End(E)$. On the other
hand, integration of superfunctions over $X$ with respect to $\mu$ gives an even $\C$-linear map: 
\beq
\nn
\int{d\mu}:\ucC(X)\rightarrow \C~~,
\eeq
which extends uniquely to an even $\End(E)$-linear map 
\beq
\nn
\left(\int{d\mu}\right)\otimes_\C \id_E:\ucC(X)\otimes_\C \End(E)\rightarrow \End(E)~~.
\eeq
In equation (\ref{completeness}) as well as below, this latter map is denoted simply by $\int{d\mu}$. 
It follows that condition (\ref{completeness}) can be viewed as a spectral
decomposition equation for the identity operator of $E$ with a
superfunction-valued spectral measure, i.e.\ a spectral decomposition taken over the
$\ucC(X)$-module $\ucC(X)\otimes_\C \End(E)$. 

Since the Berezin symbol map is injective, condition (\ref{completeness}) is
equivalent to the following (super) Fredholm equation of the first kind: \beq \nn
\int_{X} d\mu(y)\Psi(x,y)\epsilon(y)=1~~~~(x\in X)~~,  \eeq
where $\Psi(x,y)$ is the squared two point superfunction (\ref{two_point}).

When the superscalar product on $E$ is fixed, equation (\ref{completeness})
can be viewed as a constraint on those pairs $(\mu,h)$ which allow for an
integral representation of this product. Taking the
supertrace, we find a normalization condition on the epsilon superfunction:
\beq \nn \int_{X} d\mu(x) ~ \epsilon(x)=m+1-n~~, \eeq where the dimension of
$E$ is $(m+1|n)$. More details on this formula for the case when $\epsilon$ is
constant, in particular when case $m+1-n=0$, are found in section 4.4.
Equation \eqref{completeness} also allows us to establish a precise
relationship between the supertrace on $\End(E)$ and the integral over $X$:
\beq
\label{trace_identity}
\str(C)=\int_{X}d\mu(x)~\epsilon(x)\sigma(C)(x)~~.  \eeq Here $\sigma$ is the
Berezin symbol map defined by the superscalar product $(~,~)$.

Choosing a basis $s_i$, $i=0,m+n$, of $E$, we can rewrite condition
(\ref{completeness}) as a system of inhomogeneous linear integral equations for $\epsilon$:
\beq \nn \int_Xd\mu~\epsilon(x) \frac{\overline{{\hat q}(s_i)}
\hat{q}(s_j)}{\sum_{i,j=0}^{m+n}{G^{ij} \overline{{\hat q}(s_i)} {\hat
q}(s_j)}} = G_{ij}~~.  \eeq These equations, of which a subset
are independent, admit an infinity of solutions for the epsilon superfunction, so there is an
infinity of Hermitian supermetrics $h$ on $L$ which allow us to represent a given
superscalar product $(~,~)$ as an integral with respect to
$\mu_\epsilon$. Note that any such integral representation allows one to
extend the superscalar product $(~,~)$  to a Hermitian (but possibly
degenerate) pairing on the space $\Gamma(L)$ of smooth global supersections of
$L$. 

\paragraph{The relative balance condition.} The notion of balanced
metric (see \cite{Donaldson:2001aa}) 
can be extended to the case of supermanifolds, as done e.g.\ in
\cite{Catenacci:arXiv0707.4246}: 
We say that a superscalar product on $E$ is $\mu$-{\em balanced} if equation
(\ref{completeness}) 
admits a constant solution $\epsilon=\frac{m+1-n}{\mu(X)}$, i.e.\ if the following condition is satisfied:
\beq \nn \int_{X}d\mu(x)P_x\sim\id_E~~\Leftrightarrow~~\int_Xd\mu~
\frac{\overline{{\hat q}(s_i)} \hat{q}(s_j)}{\sum_{i,j=0}^{m+n}{G^{ij}
\overline{{\hat q}(s_i)} {\hat q}(s_j)}}\sim G_{ij}~.  \eeq Fixing the
proportionality constant can be sometimes subtle, because for some choices of
measure $\mu$ one can have $\mu(X)=0$. For example, the latter phenomenon occurs for some Hodge
supermanifolds with respect to the super-Liouville measure determined by their K{\"a}hler form, see e.g.\ the discussion of
the normalization of the Liouville measure in Section 4.4.

Let $\omega_h$ be the $L$-polarized K{\"a}hler form on $X$ determined by a
Hermitian superscalar product $h$ on $L$, and let $\mu_h:=\mu_{\omega_h}$ be
the Liouville measure on $X$ defined by $\omega_h$. We say that $(~,~)$ is
{\em balanced} if it is $\mu_h$-balanced. This is the case considered in
\cite{Tian-1990,Donaldson:2001aa} as mentioned in Section 2.4.

\paragraph{Remarks.} It should be stressed that, contrary to the case of
ordinary Rawnsley supercoherent states,
the supercoherent states $e_q$ do not form an overcomplete basis for $E$ in the
classical sense, because the spectral decomposition in that equation is not over $E$ but over
the auxiliary module $\ucC(X)\otimes_\C\End(E)$. In fact, $P_x$ do not even
act on the space $E$, but on the associated supermodules $E_x$ !

Note also that we have considered a number of different Hermitian pairings on
the space $\ucC(X)$ of smooth superfunctions defined on $X$. First, we have the $L^2$ pairing with
defined by the measure $\mu$: \beq
\label{mu_product}
\prec f,g \succ:=\int_{X}{d\mu {\bar f}g}~~.  \eeq Then, we have the
$L^2$ pairing defind by the measure $\mu_\epsilon=\mu \epsilon$: 
\beq
\label{epsilon_product}
\prec f,g\succ_\epsilon=\int{d\mu~ \epsilon {\bar f} g}~~.  \eeq 
Finally, the Berezin symbol space $\Sigma\subset \ucC(X)$ carries the Berezin
superscalar product: \beq
\label{induced_sp}
\prec f, g\succ_B =\vint {\bar f}\diamond g=\langle
Q(f),Q(g)\rangle_{HS}=\int_X d\mu(x) \epsilon(x) ({\bar f}\diamond g)(x)~~.
\eeq

\subsection{Generalized Toeplitz quantization}

Let us now consider the case in which the superscalar product $(~,~)$ on
$E=H^0(L)$ is determined by a measure $\mu$ on $X$ and a Hermitian form $h$ on
$L$. Since the $L^2$ Hermitian pairing induced by $(\mu,h)$ on $\Gamma(L)$ need
not be nondegenerate or positive-definite, we cannot use orthogonal
projectors from that space onto the subspace $E=H^0(L)$ of holomorphic
supersections. Instead, we return to the definition of Rawnsley supercoherent projectors,
which we extend as follows. For every $x\in X$, consider the
$\ucC_x$-supermodule $\Gamma_x:=\ucC_x\otimes_\C E$, which contains
$E_x=\ucO_x\otimes_\C E$ as a sub-supermodule via the inclusion $\ucO_x\subset
\ucC_x$. As in Section 3.1., we consider
the unique sesquilinear extension of the super Hermitian form $(~,~)$ from $E$
to the supermodule $\Gamma_x$. This is given again by eq.\ (\ref{ext}), where
now $\alpha,\beta$ belong to $\ucC_x$, and we again denote the extended form
by $\lbr~,~\rbr_x$. This non-degenerate Hermitian form 
makes $\Gamma_x$ into a Hermitian supermodule, and we can define the {\em extended
supercoherent projector} $\Pi_x\in \End_{\ucC_x}(\Gamma_x)$ by copying equation (\ref{Rw}):
\beq
\Pi_x(s)=\frac{e_q \lbr e_q,s\rbr_x}{\lbr e_q,e_q\rbr_x}\in l_x
~~~(s\in \Gamma_x)~~.
\eeq
These extended projectors are even and $\ucC_x$-linear, and their restriction
to the sub-supermodule $E_x$ recover Rawnsley's projectors:
\beq
\nn
\Pi_x|_{E_x}=P_x
\eeq
Copying equation \eqref{completeness}, we define an $\ucC(X)$-linear even
operator $\Pi$ on the $\ucC(X)$-supermodule $\Gamma(E)$ via:
\begin{equation}\label{comp}
\Pi:=\int_X d\mu(x) {\epsilon}(x)\Pi_x~~.
\end{equation}
Then it is easy to check that $\Pi$ is an idempotent operator,
i.e.\ $\Pi^2=\Pi$ and that $\im \Pi=E$.  

We are now ready to consider Toeplitz quantization.
For every smooth superfunction $f\in \ucC(X)$, define the corresponding Toeplitz
operator $T_f:=T(f)\in\End(E)$ by:
\begin{equation}\label{eqn:gen_toep}
T(f)(s)=\Pi(fs)~~\forall s\in E~~.
\end{equation}
Using \eqref{comp}, this gives:
\begin{equation}
T(f)=\int_Xd\mu(x) {\epsilon} (x) f(x) P_x~~.
\end{equation}
The underlying map $T:\ucC(X)\rightarrow \End(E)$ will be called the
generalized Toeplitz quantization of $(L,\mu,h)$. As for ordinary
manifolds, it satisfies:
\begin{equation}
 T(\bar{f})=T(f)^\dagger~~~\mbox{and}~~~T(1_X)=\id_E~.
\end{equation}
Contrary to Berezin quantization, which depends only on the
superscalar product $(~,~)$ on $E$, $T(f)$ depends essentially on the measure
$\mu_\epsilon$, which is only constrained by the completeness relation
\eqref{completeness}.

\subsection{Relation between generalized Berezin and Toeplitz quantization}

For ordinary manifolds, the generalized Berezin quantization
$Q:=\sigma^{-1}$ with respect to the superscalar product $(~,~)$ on $E$ and the
generalized Toeplitz quantization $T$ with respect to an integral
representation $(L,h,\mu)$ of this superscalar product are linked via the {\em
generalized Berezin transform}. The same holds in the case of
supermanifolds, as we will show. The map:
\begin{equation}
 \beta:=\sigma\circ T~,
\end{equation}
where $\sigma$ is the Berezin symbol map and $T$ is the Toeplitz quantization
map, is called the {\em generalized Berezin transform} and we have the integral
representation: 
\beq
\label{Berezin_tf}
\beta(f)(x)=\int_{X}{d\mu(y)~\epsilon(y)\Psi(x,y)f(y)}~~, 
\eeq 
where $\Psi$ is the squared two-point superfunction (\ref{two_point}). We now have
$T(f)=Q(\beta(f))$ and, after restricting to $\Sigma$, we find the
commutative diagram of bijections:
\beq
\nn
\begin{array}{lcc}
~~~~~\Sigma & \stackrel{T|_\Sigma}{\longrightarrow} & \End(E) \\
\beta|_\Sigma\downarrow & ~~ &\parallel \\
~~~~~\Sigma & \stackrel{Q}{\longrightarrow} &\End(E) 
\end{array}
\eeq
where $\beta$ and $T$ depend on the measure $\mu_\epsilon$ but $Q$ and
$\Sigma$ depend only on the superscalar product $(~,~)$.  Altogether, Toeplitz
quantizations associated with different integral representations of the
superscalar product $(~,~)$ on $E$ give different integral descriptions of the
Berezin quantization $Q$ defined by this product. Each Toeplitz quantization
is equivalent with $Q$ via the corresponding Berezin transform.

\paragraph{Remarks.} Let $\langle~,~\rangle_{HS}$ be the Hilbert-Schmidt
pairing on $\End(E)$ and 
$\prec~,~\succ_{\mu_\epsilon}$ be the natural super Hermitian pairing on 
$\ucC(X)$ induced by the measure $\mu_\epsilon$. As for ordinary manifolds \cite{IuliuLazaroiu:2008pk}, we have 
\begin{equation}\nn
\langle T(f),C\rangle_{HS}=\tr(T(f)^\dagger C)=\tr(T({\bar
f})C)=\int_{X}d\mu(x)\epsilon(x){\bar f}(x)\sigma(C)=\prec
f,\sigma(C)\succ_{\mu_\epsilon}~~,
\end{equation}
which shows that $T$ and $\sigma$ are adjoint to each other. It
follows immediately that $T$ is surjective, since $\sigma$ is injective and
the Berezin transform is a super Hermitian operator with image $\Sigma$.

\subsection{Changing the superscalar product in generalized Toeplitz quantization}

Let us now analyze what happens when we change the superscalar product $(~,~)$
on $E$ to $(~,~)'$ with $(s,t)':=(As,t)$. Equations \eqref{Pprime} and
\eqref{completeness} give:
\begin{equation}
\label{primecomp}
\int_{X} d\mu (x) \epsilon(x) \sigma(A^{-1})(x) P'_x=A^{-1}~~,
 \end{equation}
Using relations (\ref{productprime}),
(\ref{eprime}) and (\ref{epsilon}) we find that the epsilon superfunction of the pair $(h,(~,~)')$ is given by:
\beq
\label{rel2}
\epsilon'(x)=\epsilon(x)\sigma(A^{-1})~~,
\eeq
so (\ref{primecomp}) takes the form:
\beq
\nn
\int_{X} d\mu (x) \epsilon'(x) P'_x=A^{-1}~~.
\eeq
We can define a new Toeplitz quantization map according to: 
\beq
\nn
T'(f):=\int_{X} d\mu (x) \epsilon'(x) f(x)P'_x~~,
\eeq
which satisfies $T'(f)^\oplus=T({{\bar f}})$ as well as:
\beq
\nn
\tr(AT'(f))=\int_{X} d\mu (x) \epsilon(x)f(x)=\tr(T(f))
\eeq
and: 
\beq
\nn
T'(1_X)=A^{-1}~~.
\eeq
As on ordinary manifolds, a modified Berezin transform connects generalized
Berezin and Toeplitz quantizations with respect to the superscalar product
$(~,~)'$, cf.\ \cite{IuliuLazaroiu:2008pk}.

\subsection{Extension to powers of $L$}

The constructions of this supersection can be extended straightforwardly by
replacing the very ample super line bundle $L$ with any of its positive powers
$L^{\otimes k}$, $k\geq 1$. The new Hermitian superscalar product $(~,~)_k$ on
the supervector spaces $E_k:=H^0(L^k)$ yields new supercoherent states
$e_x^{(k)}\in E_k$ and the associated Rawnsley projectors $P_x^{(k)}$. The
latter in turn define injective Berezin symbol maps
$\sigma_k:\End(E_k)\rightarrow \ucC(X)$ whose images we denote by
$\Sigma_k$; the inverse of $\sigma_k$ after corestriction to   $\Sigma_k$ is again denoted by
$Q_k$. Note that the construction depends essentially on the sequence
$(~,~)_k$ chosen on the spaces $E_k$.

\section{Special cases: Berezin, Toeplitz and Berezin-Bergman quantization}

In this section, we first discuss the classical Berezin and Toeplitz
quantization of Hodge supermanifolds using the natural supermetric associated
to the K{\"a}hler polarization. After this general discussion, we give the
quantizations of affine and projective complex superspaces. Not
surprisingly, this is quite similar to the case of ordinary
Hodge manifolds \cite{IuliuLazaroiu:2008pk}. We also give a brief discussion of the superanalogue of
Berezin-Bergman quantization.

\subsection{Classical Berezin and Toeplitz quantization}

Given a prequantized Hodge supermanifold $(X,\omega,L,h$), we fix an integer $k_0>0$ such
that $L^{k_0}$ is very ample. For every integer $k\geq k_0$, endow $L^k$
with the Hermitian supermetric $h_k:=h^{\otimes k}$ and consider
$E_k:=H^0(L^k)$ together with the $L^2$-scalar product obtained from $h_k$ and
the Liouville measure $\mu_\omega$.

With these choices, the generalized quantization procedure yields a bijective
  symbol map $\sigma_k:\End(E_k)\rightarrow\Sigma_k\subset\ucC(X)$ and
  its inverse, the quantization map
  $Q_k:\Sigma_k\rightarrow\End(E_k)$. Moreover, we have the surjective
  Toeplitz quantization map $T_k:\ucC(X)\rightarrow\End(E_k)$. Both are
  linked by the surjective Berezin transform $\beta_k=\sigma_k\circ T_k:\ucC(X)\rightarrow \Sigma_k$ via $T_k=Q_k\circ \beta_k$. Altogether, we
  have the commutative diagram of bijections: \beq \nn
\begin{array}{rcl}
\Sigma_k & \stackrel{T_k|_\Sigma}{\longrightarrow} & \End(E_k) \\[0.1cm]
\beta_k|_\Sigma\downarrow ~& ~~~~~~~~ &~~~\parallel \\[0.1cm] \Sigma_k &
\stackrel{Q_k}{\longrightarrow} &\End(E_k)
\end{array}
\eeq

\subsection{Relations with deformation quantization and geometric quantization} 

For ordinary Hodge manifolds, it is possible to show that Toeplitz
quantization gives rise to a formal star product leading to deformation
quantizations, see
\cite{Schlichenmaier-1996aa,Schlichenmaier-1999aa,Schlichenmaier-1999bb}. Introducing
a formal Berezin transform, one can also introduce a corresponding Berezin
star product. It should be possible to extend these results to the case of
Hodge supermanifolds. For previous work on the deformation quantization of supermanifolds, see \cite{borthwick} in the cases of $U^{1|1}$ and $\C^{m|n}$ via a super-analogue of Toeplitz operators and \cite{bordemanndefquant} for split supermanifolds via a Fedosov-type procedure.

As on ordinary manifolds \cite{Souriau-1970aa,Kostant-1970aa}, one can define
a geometric quantization of a Hodge supermanifold. The prequantization
procedure goes back to \cite{Kostant:1975qe}; a (real) polarization in this
context was introduced in \cite{Tuynman:1992zm}. In the case of ordinary
manifolds, there is a clear relation between the geometric quantization of
quantizable Hermitian symmetric spaces and the Toeplitz quantization procedure
as shown in \cite{Tuynman:1987jc}. A similar relationship can be expected for
Hodge supermanifolds.

As both these points would take us too far away from the main direction of this work, we refrain from going into more detail.

\paragraph{The Berezin product or supercoherent state star product.} 
The operator product $\diamond_k:\Sigma_k\times\Sigma_k\rightarrow \Sigma_k$
introduced in section 3.2,
\begin{equation}
\label{eqn:Bproduct_k} 
f\diamond_k g\ :=\ \sigma_k(Q_k(f)Q_k(g))~,~~~f,g\in\Sigma_k~,
\end{equation} 
is also called the {\em supercoherent state star product}, since $\sigma_k(C)=\tr(C
P^{(k)}_x)$ is determined by the supercoherent states. It is associative by
definition and $(\Sigma_k,\diamond_k,\bar{~})$ is isomorphic as a
$*$-superalgebra to $(\End(E_k),\circ,\dagger)$, an isomorphism being provided
by the Berezin quantization map $Q_k$. As for ordinary manifolds, this is {\em
  not} a formal star product, cf.\ \cite{IuliuLazaroiu:2008pk}.

As an example, consider the Berezin quantization of $(\P^{m|n},\omega_{FS})$
with the prequantum super line bundle $H^k$, where $H$ is again the hyperplane
superbundle. If we normalize the homogeneous coordinates
$(Z^I)=(z^i,\zeta^\iota)=(z^0,...,z^m,\zeta^1,...,\zeta^n)$ on $\P^{m|n}$ by
demanding that $|Z|=1$, we obtain the particularly simple form
\cite{Murray:2006pi}:
\begin{equation} 
\nn
f\diamond_k
g=\sum_{I_1,...,I_k}\left(\frac{1}{k!}\der{Z^{I_1}}...
\der{Z^{I_k}}f\right)\left(\frac{1}{k!}\der{\bZ^{I_1}}...\der{\bZ^{I_k}}g\right)~.
\end{equation} 
Using the embedding $\P^{m|n}\embd \R^{m^2+n^2-1|2mn}$, one can rewrite 
this Berezin product as a finite sum of real differential operators,
resembling the first terms 
in an expansion of a formal star product, see e.g.\ \cite{Balachandran:2002jf}.

\subsection{The quantization of complex affine superspaces}

As one might expect, the Bargmann construction for the quantization of affine
space can be extended to the case of affine superspace. Again, we have to
replace the space of holomorphic supersections of the quantum super line bundle with the
space of supersections which are square integrable with respect to a weighted
version of the Liouville measure.

Consider a complex supervector space $V=V_0\oplus V_1$ of dimension $(m|n)$
over $\C$. While $V$ itself is not a supermanifold, we have the associated supermanifold
$\A_V:=(V_0,\CO_{V_0}\otimes \wedge^\bullet V_1)$ cf.\ e.g.\
\cite{Sachse:2008aa}, and in  our case
$\FC^{m|n}:=\A_V=(\FC^m,\CO_{\FC^m}[\zeta^1,...,\zeta^n])$. The structure sheaf
of $\A_V$ is freely generated by $m$ even and $n$ odd generators
$Z^I=(z^1,...,z^m,\zeta^1,...,\zeta^n)$. We denote by $B$ the algebra of
polynomials in these generators, and for any $f\in B$ we write
\begin{equation}
 f=\sum_{|\p|=\mathrm{bounded}}a_\p\chi_\p~,
\end{equation}
where $|\p|=\sum_{i=1}^{m+n} p_i$, $p_i\in \N$ for $1\leq i\leq m$ and
$p_i\in \{0,1\}$ for $m+1\leq i\leq n$. The monomials $\chi_\p$ are defined as
\begin{equation}
 \chi_\p:=Z^\p:=(z^1)^{p_1}...(z^m)^{p_m}(\zeta^1)^{p_{m+1}}...(\zeta^n)^{p_{m+n}}~.
\end{equation}
As mentioned in section 2.2, this space comes with the standard flat Hermitian
supermetric whose K{\"a}hler form is
\begin{equation}
\omega=\frac{\di}{2\pi}\left(\sum_{i=1}^m d z^i\wedge d
\bar{z}^i-i\sum_{\iota=1}^n d \zeta^\iota\wedge d
\bar{\zeta}^\iota\right)=\omega_{IL}\, d z^I\wedge d z^L
\end{equation}
and an associated Liouville measure which is given by the integral
form\footnote{Note that here and in the following $f(Z)$ specifies an
arbitrary complex superfunction, not necessarily holomorphic in the $Z^I$. In
physicists' notation, one would write $f(Z,\bZ)$.}
\begin{equation}
\begin{aligned}
 d \mu(Z)\ :=\ & \frac{1}{(2\pi)^n}|{\rm sdet}\, (\omega_{IL})|d z^1\wedge d
 \bar{z}^1\wedge...\wedge d z^m\wedge d \bar{z}^m
 id\zeta^1d\bar{\zeta}^1...id\zeta^nd\bar{\zeta}^n\\ \ =\ &
 \frac{1}{(2\pi)^{m}} d z^1\wedge d \bar{z}^1\wedge...\wedge d z^m\wedge d
 \bar{z}^m id\zeta^1d\bar{\zeta}^1...id\zeta^nd\bar{\zeta}^n~.
\end{aligned}
\end{equation}
The K{\"a}hler form is polarized with respect to the trivial super line bundle
$O:=\A_V\times \FC$. To obtain a quantum super line bundle, we endow $O$ with the
Hermitian supermetric $h$ given by
\begin{equation}
 \hat{h}(Z):=e^{-|Z|^2}~,~~~|Z|^2=\sum_{i=1}^m \bz^i z^i+i\sum_{\iota=1}^n
 \bar{\zeta}^\iota\zeta^\iota~,
\end{equation}
which corresponds to the K{\"a}hler potential $K(Z):=-\log
\hat{h}(Z)=|Z|^2$. We thus have a corresponding $L^2$-scalar product
\begin{equation}\label{Bargmannproduct}
 \langle f,g\rangle_B:=\int_{\FC^{m|n}} d \mu(Z) e^{-|Z|^2} \bar{f}(Z)g(Z)
\end{equation}
with the normalization $\langle s_0,s_0\rangle=1$, where $s_0=1$ is the unit
constant superfunction on $\FC^{m|n}$. We identify now the Bargmann space ${\cal
B}(\FC^{m|n})$ with the space of square integrable holomorphic supersections of
$O$ (which contains $B$ as a dense subset). This space carries a
representation of the Heisenberg superalgebra with $m$ even and $n$ odd pairs
of creation/annihilation operators:
\begin{equation}
\begin{aligned}
 &(\hat{a}_i^\dagger f)(Z):=z^i f(Z)~,~
 (\hat{a}_i f)(Z):=\der{z^i} f(Z)~,\\
 &(\hat{\alpha}_\iota^\dagger f)(Z):=\zeta^\iota f(Z)~,~
 (\hat{\alpha}_\iota f)(Z):=\der{\zeta^\iota} f(Z)~,
\end{aligned}
\end{equation}
or, summarizing them according to
$(\hat{A}_I)=(\hat{a}_i,\hat{\alpha}_\iota)$:
\begin{equation}\nn
 (\hat{A}_I^\dagger f)(Z):=Z^I f(Z)~,~~~ (\hat{A}_I f)(Z):=\der{Z^I} f(Z)~.
\end{equation}
These operators satisfy the commutation relations\footnote{The factor of $\di$
is necessary to match our conventions for complex conjugation of objects of
odd parity:
$(\hat{\alpha}_\alpha\hat{\alpha}_\beta)^\dagger=-\hat{\alpha}^\dagger_\beta\hat{\alpha}^\dagger_\alpha$.}
\begin{equation}\label{eq:CommutationRelationsAffine}
 [\hat{a}_i,\hat{a}_j^\dagger]:=[\hat{a}_i,\hat{a}_j^\dagger]_-=\delta_{ij}~,~~~\{\hat{\alpha}_\iota,\hat{\alpha}_\gamma^\dagger\}:=[\hat{\alpha}_\iota,\hat{\alpha}_\gamma^\dagger]_+=\di\delta_{\iota\gamma}~,
\end{equation}
or, using the supercommutator $\lsc~,~\rsc$:
\begin{equation}
 \lsc\hat{A}_I,\hat{A}_J^\dagger\rsc=\di^{\tilde{I}\tilde{J}}\delta_{IJ}~.
\end{equation}
We normalize the vacuum vector in ${\cal B}(\FC^{m|n})$ to the constant unit
function $|0\rangle:=1$, and setting $\langle 0|0 \rangle_B=1$ yields together
with the commutation relations \eqref{eq:CommutationRelationsAffine} a
Hermitian superscalar product $\langle~|~\rangle_B$ on ${\cal B}$.  The
normalized occupation vectors are given by:
\begin{equation}\label{occvecs}
|\p\rangle=\frac{1}{\sqrt{\p!}}\chi_\p=\frac{({\hat
 A}^\dagger)^\p}{\sqrt{\p!}}|0\rangle~~~\mbox{with}~~~||\chi_\p||^2_B=(i)^{(\sum_{\iota=1}^n
 p_\iota) {\rm mod}\,2} \p!=(i)^{\widetilde{|\p\rangle}} \p!~,
\end{equation}
where $\p!:=p_0!\ldots p_n!$. Defining the number operators
$\hat{N}_I:=(-i)^{\tilde{I}}\hat{A}^\dagger_I\hat{A}_I$, we have ${\hat
N}_I|\p\rangle=p_I|\p\rangle$. The total number operator
\begin{equation}
 \hat{N}:=\sum_{I=1}^{m+n} \hat{N}_I=\sum_{i=1}^m
 \hat{a}^\dagger_i\hat{a}_i-i\sum_{\iota=1}^n\hat{\alpha}^\dagger_\iota\hat{\alpha}_\iota
\end{equation}
allows us to introduce the decomposition ${\cal
B}:=\overline{\oplus}_{k=0}^\infty B_k$ with $B_k=\ker({\hat N}-k)$.

We define the supercoherent vectors with respect to $q=s_0(z)=1\in \CO_z$ and
this definition yields the usual super Glauber vectors:
\begin{equation}\label{eqn:glauber}
\begin{aligned}
 |Z\rangle=e^{\sum_{I=1}^{m+n}i^{\tilde{I}}{\bar Z}^I{\hat
    A}_I^\dagger}|0\rangle&=e^{\sum_{i=1}^{m}\bz^i\hat{a}_i^\dagger+i\sum_{\iota=1}^n\bar{\zeta}^\iota\hat{\alpha}_\iota^\dagger}|0\rangle~,\\
    |Z\rangle&=\sum_{\p}(i)^{\widetilde{|\p\rangle}}{\frac{\bZ^\p}{\sqrt{\p!}}|\p\rangle}~~,\\\hat{A}_I|Z\rangle=\bZ^I|Z\rangle~~&,~~\langle
    Z_1|Z_2\rangle_B=e^{(Z_2,Z_1)}~,
\end{aligned}
\end{equation}
where $\bZ^\p=\bz_1^{p_1}\ldots
\bz_n^{p_m}\bar{\zeta}^{p_{m+1}}_1...\bar{\zeta}^{p_{m+n}}_n$ and
$(Z_1,Z_2):=\sum_{i=1}^m \bz^i_1 z^i_2+i\sum_{\iota=1}^n
\bar{\zeta}^\iota_1\zeta^\iota_2$ denotes the superscalar product on
$\FC^{m|n}$. We have as usual \beq \nn f(Z)=~\langle
Z|f\rangle_B~~~\mbox{for}~~~f\in {\cal B}~~.  \eeq The reproducing kernel is
the super Bergman kernel: \beq
\label{eqn:bergmankernel}
K_B(Z_1,Z_2)=\frac{\langle Z_1|Z_2\rangle}{\sqrt{\langle Z_1|Z_1\rangle \langle
    Z_2|Z_2\rangle}}=e^{-\frac{1}{2}(|Z_1|^2+|Z_2|^2)+(Z_2,Z_1)}~~.
\eeq The Rawnsley projector is given by
\begin{equation}
 P_Z=\frac{1}{\langle Z| Z\rangle_B}|Z\rangle \langle Z|_B=e^{-|Z|^2}
|Z\rangle \langle Z|_B
\end{equation}
with constant epsilon superfunction $\epsilon_{\C^{m|n}}(Z)={\hat h}(Z)\langle
Z|Z\rangle_B=1$ and decomposition of the identity $\int_{\FC^{m|n}} d\mu(Z)
P_Z=\unit_{\cal B}$~.

\paragraph{Toeplitz quantization of $\A_V$.} The Toeplitz quantization of $f\in \ucC(\FC^{m|n})$ is given by:
\beq
\label{plane_T}
T(f)=\int_{\FC^{m|n}}d\mu(Z) f(Z)P_Z=\int_{\FC^{m|n}} d\mu(Z) e^{-|Z|^2}
f(Z)|Z\rangle \langle Z|_B~~.  \eeq In particular, we have $T(Z^I)={\hat
A}_I^\dagger$ and $T({\bar Z}^I)={\hat A}_I$.  When $f$ is a polynomial in $Z$
and $\bZ$, (\ref{plane_T}) obviously reduces to the anti-Wick prescription:
\beq \nn T(f(Z,\bZ))=\vdots f({\hat A}^\dagger, {\hat A})\vdots~~, \eeq where
the triple dots indicate antinormal ordering. In this case, $T$ is not
surjective due to the infinite-dimensionality of the Bargmann space.

\paragraph{Berezin quantization of $\A_V$.} The Berezin symbol map is easily
extended as well. 
It is defined on the algebra ${\cal L}({\cal B})$ of bounded operators in the Bargmann space and maps them into $\ucC(\C^{m|n})$ as follows: \beq \nn \sigma(C)(Z)=e^{-|Z|^2}\langle
Z|C|Z\rangle_B~~.  \eeq The Berezin transform $\beta(f)=\sigma\circ T$ is
given by: \beq \nn \beta(f)(Z_1)=\int_{\FC^{m|n}} d\mu(Z_2)
f(Z_2)e^{-|Z_1-Z_2|^2}~~.  \eeq

The symbol map gives rise to the Berezin quantization map $Q:\Sigma\rightarrow
{\cal L}({\cal B})$, where $\Sigma\subset\ucC(\C^{m|n})$ is the
image of $\sigma$. We have $Q(Z^I)={\hat A}_I^\dagger$ and $Q({\bar
Z}^I)={\hat A}_I$. For a polynomial superfunction $f(Z,\bZ)$, we find: \beq \nn
Q(f)=:f({\hat A}^\dagger, {\hat A}):~~, \eeq where the double dots indicate
normal ordering. Hence both quantization prescriptions send $Z^I$ into ${\hat
A}_I^\dagger$ and ${\bar Z}^I$ into ${\hat A}_I$, but Toeplitz quantization
corresponds to anti-Wick ordering, while Berezin quantization corresponds to
Wick ordering.

\paragraph{Restricted supercoherent vectors.} For later use, 
consider the expansion of Glau\-ber's supercoherent vectors $|Z\rangle$ 
in components $|Z,k\rangle$ of fixed total particle number $k$, i.e.\ $\hat{N}|Z,k\rangle=k$:
\beq \nn |Z\rangle=\sum_{k=0}^\infty |Z,k
\rangle~,~~~|Z,k\rangle:=\frac{1}{k!}\left(\sum_{i=1}^{m}\bz^i\hat{a}_i^\dagger+
i\sum_{\iota=1}^n\bar{\zeta}^\iota\hat{\alpha}_\iota^\dagger\right)^k|0\rangle~.
\eeq We note for future reference that $\langle Z,k|Z,k\rangle_B=\frac{1}{
k!}|Z|^{2k}$ and ${\hat A}_I |Z,k\rangle=\bZ^I|Z,k-1\rangle$. Note furthermore
that $|\lambda Z,k\rangle={\bar \lambda}^k|Z,k\rangle$ for any $\lambda \in
\C$, and therefore the ray $\C^*|Z,k\rangle$ depends only on the image $[Z]$
of $Z$ in the projective superspace $\P^{m-1|n}$.

\subsection{The quantization of complex projective superspaces}

It is now easy to carry out the quantization of complex projective superspaces. For
earlier discussions of these spaces relying on group
theoretic methods, see \cite{Ivanov:2003qq,Murray:2006pi}. Another possible approach would be to extend the techniques of \cite{bordemanncpn1,bordemanncpn2} to the supercase.

Consider the supermanifold $\P^{m|n}$ as introduced in section 2.2 with
homogeneous supercoordinates $Z^I=(z^0,...,z^m,\zeta^1,...,\zeta^n)$. As a quantum
super line bundle, we take the super hyperplane bundle $H:=O(1)$, which is very ample. The
space of supersections $H^0(H^k)$ is the space of homogeneous polynomials of degree
$k$ and can thus be identified with $B_k\in {\cal B}$, where ${\cal B}$ is the
Bargmann space used in the quantization of $\FC^{m+1|n}$. Notice that:
\begin{equation}
\begin{aligned}
 {\rm dim} B_k&=(b_k^0|b_k^1)~,\\ b_k^0=\sum_{i=0}^{[{\rm min}\{k,n\}/2]} &\frac{(m+1+(k-2i))!}{(m+1)!(k-2i)!}\frac{n!}{(n-2i)!(2i)!}~,\\
 b_k^1=\sum_{i=0}^{[({\rm min}\{k,n\}-1)/2]}
 &\frac{(m+1+(k-(2i+1))!}{(m+1)!(k-(2i+1))!}\frac{n!}{(n-(2i+1))!(2i+1)!}~,
\end{aligned}
\end{equation}
where $b_k^0$ and $b_k^1$ are the even and odd dimensions of $B_k$,
respectively, and $[..]$ denotes taking the integral part. The first
factor in the sums corresponds to the symmetrized even homogeneous coordinates
$z^i$, while the second factor represents the antisymmetrized odd homogeneous
coordinates $\zeta^{\iota}$.

We endow the hyperplane superbundle $H$ with the Hermitian supermetric given by 
\beq
\label{eqn:hypermetric}
h_{FS}([Z])(Z^I,Z^I)=\frac{|Z^I|^2}{|Z|^2}~,
\eeq
which is associated to the following K{\"a}hler supermetric on $\P^{m|n}$:
\beq
\label{eqn:hyperform}
\omega_{FS}([Z])=\frac{i}{2\pi}\partial {\bar \partial}\log
|Z|^2=\frac{i}{2\pi}\partial {\bar \partial}\log
(1+|Z_0|^2)~~, \eeq cf.\ Section 2.2. Let us be more explicit and restrict to
the patch $U_0$ for which $z^0\neq 0$ with local coordinates
$(Z_0^I)=(z_0^i,\zeta_0^\iota)$, $I=1,...,m+n$, where $z_0^i=\frac{z^i}{z^0}$
and $\zeta_0^\iota=\frac{\zeta^\iota}{z^0}$. On this patch, the K{\"a}hler
form reads as
\begin{equation}
 \omega_{FS}|_{U_0}=\omega_{IL}\,d Z_0^I\wedge d Z_0^L
\end{equation}
with
\begin{equation}\nn
 \omega_{IL}=\left(\begin{array}{c|c} \omega_{il} & \omega_{i\lambda} \\
                  \hline
                \omega_{\iota l} & \omega_{\iota\lambda}
                 \end{array}\right)=\frac{\di}{2\pi(1+|Z_0|^2)^2}\left(\begin{array}{c|c} \delta^{il}(1+|Z_0|^2)-\bz^i_0z^l_0 & -\di \bz^i_0 \zeta^\lambda_0\\
                  \hline\\[-0.5cm]
                -\di \bar{\zeta}^\iota_0 z^l_0 & \di\delta^{\iota\lambda}(1+|Z_0|^2)-\bar{\zeta}^\iota_0\zeta^\lambda_0
                 \end{array}\right)~.
\end{equation}
The corresponding Liouville measure $d \mu(Z)$ is given in the coordinates on
the patch $U_0$ as
\begin{equation}\label{eq:LiouvilleMeasure}
(2\pi)^n|{\rm sdet}\, (\omega_{IL})|d z^1_0\wedge d \bar{z}^1_0\wedge...\wedge
d z^m_0\wedge d \bar{z}^m_0 \di d\zeta^1_0d\bar{\zeta}^1_0...\di
d\zeta^n_0d\bar{\zeta}^n_0~,
\end{equation}
where
\begin{equation}
 |{\rm sdet}\, (\omega_{IL})|:=\frac{\det((\omega_{il})-(\omega_{i\lambda})(\omega_{\iota\lambda})^{-1}(\omega_{\iota l}))}{\det(\omega_{\iota\lambda})} =(2\pi)^{n-m}(1+|Z_0|^2)^{n-m-1}~.
\end{equation}
Note that the volume of $\P^{m|n}$ vanishes, if $(n-m-1)\geq 0$ because of the
Berezin integration in the measure. Otherwise, we can use the formula
\begin{equation}\label{eqn:FermExp}
 \frac{1}{(1+\sum_i \bz^i_0 z^i_0+i
 \sum_\iota\bar{\zeta}_0^\iota\zeta_0^\iota)^g}=\sum_{\ell=0}^n
 (-1)^\ell\frac{(g-1+\ell)!}{(g-1)!\ell!}\frac{1}{(1+\sum_i \bz^i_0
 z^i_0)^{g+\ell}}\left(i
 \sum_\iota\bar{\zeta}_0^\iota\zeta_0^\iota\right)^\ell
\end{equation}
and arrived at the closed expression
\begin{equation}
 \vol_{\omega_{FS}}(\P^{m|n})=\frac{1}{m!}\frac{m!}{(m-n)!}~,
\end{equation}
and in particular, we have $\vol_{\omega_{FS}}(\P^{m|0})=\frac{1}{m!}$.

The supermetric on the hyperplane superbundle $H$ extends to the tensor product
supermetric $h_{FS}^k:=h_{FS}^{\otimes k}$, which satisfies: \beq
\label{eqn:hypertensor}
h_{FS}^k([Z])(S([Z]),S([Z]))=\frac{|s(Z)|^2}{|Z|^{2k}} \eeq for all
supersections $S\in H^0(H^k)$ and their corresponding $s\in B_k$. The space
$H^0(H^k)\cong B_k$ carries the associated $L^2$-product: \beq
\label{eqn:hyperproduct}
\langle s_1,s_2\rangle_k:=\langle S_1,S_2\rangle^{h^k_{FS}}_k=\int_{\P^{m|n}}
d \mu(z)~h^k_{FS}(S_1,S_2)~~.  \eeq Note that the monomials $\chi_\p$ with
$|\p|=k$ provide an orthogonal but not orthonormal basis of $B_k$ with respect
to the superscalar product \eqref{eqn:hyperproduct}. Using formula
\eqref{eqn:FermExp}, we easily verify that
\begin{equation}\label{products}
\langle s, t\rangle_k= \frac{1}{(m-n+k)!}\langle s,t\rangle_B~~~\forall s,t\in
B_k~~.
\end{equation}

The quantization of $\P^{m|n}$ proceeds now in a straightforward manner. The
supercoherent states of the quantum super line bundle $H^k$ are the Glauber
supercoherent states restricted at level $k$ and from these, we construct the
supercoherent projectors \beq
\label{eqn:perelomov}
P^{(k)}_{[Z]}:=\frac{|Z,k\rangle{}\langle Z,k|_B}{\langle Z,k|Z,k\rangle_B}~.
\eeq
The overcompleteness relation takes the form \beq
\label{Pcompleteness}
(b_k^0-b_k^1)\int_{\P^{m|n}}d\mu([Z]) P^{(k)}_{[Z]}=\vol(\P^{m|n})P_k~~, \eeq
where $P_k$ is the orthoprojector on $B_k$ in ${\cal B}(\FC^{m|n})$. The
normalization is obtained by taking the supertrace of both sides. Note that
interestingly whenever $m<n$ and thus $\vol(\P^{m|n})=0$, we also have
$b_k^0-b_k^1=0$ for $k>0$ as one can show e.g.\ by complete
induction. Therefore, this normalization condition does not give any
additional constraint in these cases. Alternatively, we can calculate the
ordinary trace. While ${\rm str}\,P^{(k)}_{[Z]}=1$ and ${\rm
str}\,P_k=b_0^k-b_1^k$, we have
\begin{equation}
 {\rm tr}\,(P^{(k)}_{[Z]})=\left(\frac{\sum_{i=0}^m \bz^i z^i-i
 \sum_{\iota=1}^n\bar{\zeta}^\iota\zeta^\iota}{\sum_{i=0}^m \bz^i z^i+i
 \sum_{\iota=1}^n\bar{\zeta}^\iota\zeta^\iota}\right)^k~,~~~{\rm
 tr}\,P_k=b_0^k+b_1^k~.
\end{equation}
The expression
\begin{equation}\label{def:TildedVolume}
 \vol_{\tr}(\P^{m|n}):=\int_{\P^{m|n}}d\mu([Z]) \left(\frac{\sum_{i=0}^m \bz^i
 z^i-i \sum_{\iota=1}^n\bar{\zeta}^\iota\zeta^\iota}{\sum_{i=0}^m \bz^i z^i+i
 \sum_{\iota=1}^n\bar{\zeta}^\iota\zeta^\iota}\right)^k
\end{equation}
is clearly non-vanishing and can easily be evaluated in every concrete
case. Our new normalization of the overcompleteness relation \eqref{Pcompleteness}
now reads as
\begin{equation}
 \frac{b_k^0+b_k^1}{\vol_{\tr}(\P^{m|n})}\int_{\P^{m|n}}d\mu([Z])
 P^{(k)}_{[Z]}=P_k~.
\end{equation}

We will restrict ourselves to superfunctions on $\P^{m|n}$ of the form: \beq
\label{f}
f_{{\cal I}{\cal J}}([Z]):=\frac{\bZ^{\cal I} Z^{\cal
  J}}{|Z|^{2k}}:=\frac{({\bar z}^0)^{{\cal I}_0}\ldots (\bar{\zeta}^n)^{{\cal
  I}_{m+n}}(z^0)^{{\cal J}_0}\ldots (\zeta^n)^{{\cal J}_{m+n}}}{|Z|^{2k}}~~,
  \eeq $f_{{\cal IJ}}[z]\in \ucC(\P^{m|n})$, where ${\cal I}=({\cal
  I}_0\ldots {\cal I}_{m+n}),{\cal J}=({\cal J}_0\ldots {\cal J}_{m+n})$ with
  ${\cal I}_L,{\cal J}_L\in \N$ for $L\leq m$ and ${\cal I}_L,{\cal
  J}_L\in\{0,1\}$ for $L>m$ and $|{\cal I}|=|{\cal J}|=k$ and where we set
  $f_{{\cal IJ}}=1$ for $m=n=0$. Furthermore,
  we can decompose ${\cal S}(\P^{m|n})$ into the subsets ${\cal
  S}_k(\P^{m|n})$, which are spanned by the superfunctions $f_{{\cal IJ}}$
  with $|{\cal I}|=|{\cal J}|=k$; note that ${\cal S}_0(\P^{m|n})=\C$. For any
  $L=0\ldots m+n$, let $\Delta_L\in \N^{m+n+1}$ be given by
  $\Delta_L(I)=\delta_{IL}$. The obvious relation: \beq \nn f_{\cal
  IJ}=\sum_{L=0}^{m+n} f_{{\cal I}+\Delta_L, {\cal J}+\Delta_L} \eeq shows
  that ${\cal S}_k(\P^{m|n})\subset {\cal S}_{k+1}(\P^{m|n})$ for all $k\geq
  0$, so that ${\cal S}(\P^{m|n})=\cup_{k=0}^\infty{\cal S}_k(\P^{m|n})$ is a
  filtered $*$-algebra generated by the elements $f_{IJ}=\frac{{\bar
  Z}_IZ_J}{|Z|^2}\in {\cal S}_1(\P^{m|n})$.

Note that the space ${\cal S}(\P^{m|n})$ forms a good approximation to ${\cal
C}^\infty(\P^{m|n}$): Since $\P^{m|n}$ is a split supermanifold, any
superfunction $f\in \ucC(\P^{m|n})$ allows for a globally valid
expansion of the form
\begin{equation}\label{decomposition}
 f(Z)=\sum_{|\cal A|+|\cal C|=|\cal B|+|\cal D|} f_{\cal A\cal B\cal C\cal
 D}(z) \frac{\zeta^{\cal A} \bar{\zeta}^{\cal B}z^{\cal C}\bz^{\cal
 D}}{|Z|^{|\cal A|+|\cal B|+|\cal C|+|\cal D|}}
\end{equation}
with multi-indices ${\cal A},{\cal B}$ and coefficient superfunctions $f_{\cal
A\cal B\cal C\cal D}(z)\in \ucC(\P^{m|0})$; the latter are well
approximated, as ${\cal S}(\P^{m|0})$, which is contained in ${\cal
S}(\P^{m|n})$, is dense in $(\CC^\infty(\P^{m}),||~||_\infty)$. The latter is
true due to the Stone-Weierstra\ss{} theorem, cf.\
\cite{IuliuLazaroiu:2008pk}. To define an orthoprojector $\pi_k$ onto ${\cal
S}_k(\P^{m|n})$, we cannot rely on an $L^2$-scalar product on $\P^{m|n}$. We
can, however, project each coefficient superfunction $f_{\cal A\cal B\cal
C\cal D}$ with $k-|{\cal A}|-|{\cal C}|\leq 0$ onto ${\cal S}_{k-|{\cal
A}|-|{\cal C}|}(\P^{m|0})$ using the ordinary $L^2$-orthoprojector on ${\cal
C}^\infty(\P^m)$ and plug these back into the expansion
\eqref{decomposition}. This clearly yields an element of ${\cal
S}_k(\P^{m|n})$.

\paragraph{Toeplitz quantization of $\P^{m|n}$.} We define the Toeplitz
quantization map 
$T_k:\CC^\infty(\P^{m|n},\FC)\rightarrow \sEnd(B_k)$ according to
\begin{equation}
T_k(f)=\frac{b_k^0+b_k^1}{\vol_{\tr}(\P^{m|n})}\int_{\P^{m|n}} d \mu([Z])
f([Z])P^{(k)}_{[Z]}~~.
\end{equation}

Due to ${\hat A}_I |Z,k\rangle=\bar{Z^I}|Z,k-1\rangle$, we find: \beq
\begin{aligned}
T_k(f_{\cal IJ})&= \frac{b_k^0+b_k^1}{\vol_{\tr}(\P^{m|n})}\int_{\P^{m|n}}
  d\mu([Z]) \frac{{\hat A}^{\cal I}|Z,k+d\rangle\langle Z,k+d|_B ({\hat
  A}^\dagger)^{\cal J}}{|Z|^{2(k+m)}}\\&= \frac{b_k^0+b_k^1}{\vol_{\tr}(\P^{m|n})}
  {\hat A}^{\cal I} P_{k+m}({\hat A}^\dagger)^{\cal J}~~,
\end{aligned}
\eeq and thus the map $T_k(f)$ is surjective. As a special case, we have: \beq
\nn T_k(f_{IJ})=\frac{b_1^0+b_1^1}{\vol_{\tr}(\P^{m|n})}{\hat A}_I{\hat
A}_J^\dagger~~.  \eeq

\paragraph{Berezin quantization of $\P^{m|n}$.} The Berezin symbol map $\sigma_k:\End(B_k)\rightarrow {\cal
  C}^\infty(\P^{m|n},$ $ \C)$ takes the form: \beq \nn
\sigma_k(C)([Z])=\frac{\langle Z , k|C|Z, k\rangle}{ \langle Z, k|Z,
k\rangle}~~~~\forall C\in \End(B_k)~~.  \eeq This map is injective, and we can
define an inverse on $\Sigma_k:=\im \sigma_k$ which yields the Berezin
quantization map $Q_k:\Sigma_k(\P^{m|n})\rightarrow
\End(B_k)$, which is a linear isomorphism. Under quantization, the superfunctions (\ref{f}) are mapped to:
\beq \nn Q_k(f_{\cal IJ})=\frac{1}{k!}P_k({\hat A}^\dagger)^{\cal I} {\hat
A}^{\cal J}P_k~, \eeq and we have in particular: \beq \nn
Q_k(f_{IJ})=\hat{A}_J^\dagger \hat{A}_I~~.  \eeq Notice that the operators
${\hat f}_{\cal IJ}=P_k({\hat A}^\dagger)^{\cal I} {\hat A}^{\cal J}P_k$ with
$|{\cal I}|=|{\cal J}|=k$ provide a basis for $\End(B_k)$, and thus the image
$\Sigma_k(\P^{m|n})$ can be identified with ${\cal S}_k(\P^{m|n})$. Therefore,
$\Sigma_k(\P^{m|n})$ provides a weakly exhaustive filtration of $( {\cal
C}^\infty(\P^{m|n}),||~||^\circ_\infty)$: \beq \nn \overline{\cup_{k=1}^\infty
\Sigma_k(\P^{m|n})}=\ucC(\P^{m|n})~~.  \eeq The Berezin transform
$\beta_k:\ucC(\P^{m|n})\rightarrow \Sigma_k(\P^{m|n})$ is here defined as:
\beq \nn
\beta_k(f)([Z])=\sigma_k(T_k(f))=\frac{b_k^0+b_k^1}{\vol_{\tr}(\P^{m|n})}
\int_{\P^{m|n}}d\mu([Y])\left(\frac{|(Y,k|Z,k)|}{(Y,k|Y,k)(Z,k|Z,k)}\right)^{2k}~~.
\eeq As in the quantization of affine space, Berezin and Toeplitz
quantizations use Wick and anti-Wick orderings, respectively.

\paragraph{Remarks.} As the space $\P^{m|n}$ is the coset space
$U(m+1|n)/(U(1|0)\times U(m|n))$, 
the Rawnsley supercoherent states can be identified with the Perelomov
supercoherent states. Rather obviously, 
the spaces $B_k$ and ${\cal B}_k$ form irreducible representations of the
supergroup $U(m+1|n)$. 
For further details on the group theoretic aspects of Berezin-quantized $\P^{m|n}$, see \cite{Murray:2006pi}.

\subsection{Berezin-Bergman quantization}

In \cite{Saemann:2006gf}, a quantization prescription was proposed for
projective algebraic varieties, which relied on their embedding into
projective space. More explicitly, the idea was to use the identification of
supersections of the quantum bundle $H^0(H^k)$ on $\P^m$ and the Hilbert space
$B_k$ in the quantization of the embedded variety $X$ by factoring out an
ideal. The zero locus conditions $f_i=0$ defining $X\subset \P^m$ reducing the
space $H^0(H^k)$ would go over into conditions $\hat{f}_i|\mu\rangle=0$ for
all $|\mu\rangle\in B_k$. As shown in \cite{IuliuLazaroiu:2008pk}, this {\em
Berezin-Bergman quantization} corresponds to a generalized Berezin
quantization. A similar construction can also be performed
in the case of Hodge supermanifolds and we outline this construction in
the following.

We start from a polarized complex supermanifold $(X,L)$ with very ample 
super line bundle $L$ and $\dim_\FC H^0(L)=(m|n)$. The homogeneous coordinate ring
of $X$ is (bi-)graded:
$R(X,L)=\oplus_{k=0}^\infty{H^0(L^k)}=:\oplus_{k=0}^\infty E_k$, and we have
an isomorphism of graded algebras
$\phi:R\stackrel{\sim}{\rightarrow}B/I$. Here, $B$ is the graded symmetric
algebra $B=\oplus_{k=0}^\infty E_1^{\odot k}$ and $I=\oplus_{k=0}^\infty I_k$
is a graded ideal in $B$. The Kodaira superembedding theorem \cite{MR1032867},
gives a superembedding defined by $L$ in which
$X$ is presented as a projective algebraic supervariety in $\P^{m|n}$ with
vanishing ideal $I$. We have
\begin{equation}
 I_k\subset B_k~~~\mbox{and}~~~E_k\simeq B_k/I_k~.
\end{equation}

At every level $k$, one has two natural choices for introducing a super Hermitian
pairing on $H^0(L^k)$. The first is to take the usual $L^2$-product:
\beq \nn \langle s,t\rangle_k=\int_{X}d\mu_\omega h^{\otimes k}(s,t)~~, \eeq
while the second one is induced from $B$ as follows:
\begin{equation}
 ( s_1\odot \ldots s_k,t_1\odot \ldots
  t_l)_B=\frac{1}{k!}\delta_{k,l}\sum_{\sigma\in
  S_k}{\epsilon(\sigma,t_1\ldots t_k)(s_1,t_{\sigma(1)})_1\ldots (s_k,t_{\sigma(k)})_1}~~.
\end{equation}
Here $(~,~)_1$ is the superscalar product on $E_1$, $S_k$ is the symmetric
group on $k$ letters, $s_i,t_i\in E_1$ and $\epsilon(\sigma,t_1\ldots t_k)$ is
the Koszul sign in the graded symmetric product. Notice that
the second choice is actually the restriction of \eqref{Bargmannproduct} to
$B_k$.

Let $I_k^\perp:=\{s\in B_k|(s,t)_B=0~~\forall t\in I_k\}$ and notice that we 
can identify this space with $E_k$. First, we can identify $B_k$ with
$H^0(H^k)$, where $H$ is the super hyperplane bundle over $\P^{m|n}$. Then we have a
restriction $i_k^*:H^0(H^k)=B_k\rightarrow H^0(L^k)=E_k$. As $I_k=\ker i^*_k$
and since $i_k^*$ is surjective, we have an isomorphism
$\phi_k:=E_k\rightarrow I_k^\perp\simeq B_k/I_k$. Then we define a superscalar
product $(~,~)_k$ on $E_k$ via: \beq
\label{induced_product}
(s,t)_k:=\alpha_k( \phi_k(s),\phi_k(t))_B~~.  
\eeq 
For ordinary manifolds, the choice of the normalization constants $\alpha_k$ depended on the volumes
of $X$ and $\P^{m|n}$ as well as the dimensions of $B_k$ and $E_k$. In the
supercase, one again has to introduce the trace volume $\vol_\tr$ if the classical supervolume
of $X$ or $\P^{m|n}$ vanishes. We can then impose the usual condition for a
``good'' quantization: that under generalized Berezin quantization, the unit
superfunction $1_X$ is mapped into the unit operator $\unit$ on $E_k$.

\paragraph{Definition.} The {\em Berezin-Bergman quantization} of $(X,L)$
determined by the superscalar product $(~,~)_1$ on $H^0(L)$ is the generalized
Berezin quantization performed with respect to the sequence of superscalar products
$(~,~)_k$ on $H^0(L^k)$ defined in (\ref{induced_product}).

\paragraph{Remarks.} As the vanishing ideal $I$ is zero in the case of
  $\P^{m|n}$, 
Berezin-Bergman quantization here corresponds to ordinary Berezin quantization.

If $I$ is generated by $p$ homogeneous polynomials $F_1\ldots F_p$ of degrees
at least two, then we have \beq \nn I^\perp=\cap_{l=1}^p{\ker {\bar F}_l({\hat
A}^\dagger)}~~, \eeq where ${\bar F}$ is the polynomial in $Z_I$ obtained by
conjugating all {\em coefficients} of $F$ as in the case of ordinary
manifolds, cf.\ \cite{IuliuLazaroiu:2008pk}.

\section{Regularizing supersymmetric quantum field theories}

As stated in the introduction, one of the major applications of
Berezin-quantized manifolds in physics is the regularization of path integrals
and the numerical treatment of quantum field theories. In this section, we
extend the definition of fuzzy superscalar field theories, i.e.\ superscalar field
theories defined on Berezin-quantized Hodge manifolds, to some supersymmetric
cases. For the earliest work in this direction, see \cite{Grosse:1995pr}; our
discussion will follow along similar lines as those proposed in
\cite{Klimcik:1999pr}.

It should be clear that an exhaustive discussion of supersymmetric superscalar
field theories on quantized supermanifolds, which, as we will see, requires
that they admit a superfield description, cannot possibly\footnote{cf.\
e.g.\ \cite{Hubsch:1998ps} just for the case of two dimensions} be performed
within this work. We will therefore restrict our discussion to giving an
example in more detail: the ${\cal N}=(2,2)$ supersymmetric sigma model on
Berezin-quantized $\P^{1|2}$. We will also comment on its topological twist,
which can, in principle, be defined on an arbitrary Riemann surface. These
theories are particularly interesting, as they serve as the basic building
blocks for string theories.

\subsection{Fuzzy scalar field theories}

Classical (real) scalar field theory on a K{\"a}hler manifold $(X,\omega)$ is
usually given by an action functional of the form
\begin{equation}\label{eq:ScalarAction}
 S[\phi]=\frac{1}{{\rm vol}_\omega(X)}\int_X\frac{\omega^n}{n!}~\big(\phi
 \Delta\phi+V(\phi)\big)~,~~~\phi\in\CC^\infty(X,\FR)~,
\end{equation}
where $\Delta$ is the Laplace operator on $(X,\omega)$ and $V(\phi)$, the {\em
potential}, is a polynomial in $\phi$ with real coefficients. To study a
similar field theory on a quantized manifold, we need a quantization of the
classical Laplace operator. Two such quantizations are possible and we will
briefly review them below. For a more detailed discussion, see
\cite{IuliuLazaroiu:2008pk}.

In the following, consider a quantized Hodge manifold $(X,\omega,E_k)$ with
symbol space $\Sigma_k=\sigma(\End(E_k))$. In general, there are two ways of
defining a quantum analogue to an operator $\CD:\CC^\infty(X)\rightarrow
\CC^\infty(X)$. First, we define by $\CD_k$ a truncated map
$\Sigma_k\rightarrow \Sigma_k$ as:
\begin{equation}
 \CD_k:=\pi_k\circ \CD|_{\Sigma_k}~,
\end{equation}
where $\pi_k$ is the orthoprojector with respect to the ordinary $L^2$-scalar
product on $(X,\omega)$. The Berezin push $\CD_k^B$ of an operator $\CD$ is
then defined as the following map $\CD^B_k:\End(E_k)\rightarrow \End(E_k)$:
\begin{equation}
 \CD_k^B:=Q_k\circ \CD_k\circ \sigma_k~.
\end{equation}
Roughly speaking, the Berezin push of an operator acts as the corresponding
operator in the continuum (up to truncations), and we have in particular in
the case of the identity operator $\CD(f)=f$ for all $f\in\CC^\infty(X)$ the
quantization $\CD_k^B(\hat{f})=\hat{f}$ for all
$\hat{f}\in\End(E_k)$. Hermitian operators with respect to the natural
$L^2$-norm on $(X,\omega)$ are, however, not mapped into Hermitian operators
with respect to the natural Hilbert-Schmidt norm on $\End(E_k)$.

Alternatively, we can define the {\em Berezin-Toeplitz lift} of an operator
$\CD:\CC^\infty(X)\rightarrow \CC^\infty(X)$ as
\begin{equation}
 \hat{\CD}_k:=T_k\circ M_{\frac{1}{\epsilon_k}}\circ\CD\circ \sigma_k~,
\end{equation}
$\hat{\CD}:\End(E_k)\rightarrow \End(E_k)$, where $M_{\alpha}(f):=\alpha f$,
$\alpha,f\in\CC^\infty(X)$ is the multiplication operator. This operator will
not map the identity on $\CC^\infty$ onto the identity on $\End(E_k)$, but the
hermiticity of operators is preserved under quantization.

As a side remark, note that the definition of a Berezin push and a
Berezin-Toeplitz lift of operators readily extends to quantized
supermanifolds. For the Berezin push, one can use the orthoprojector $\pi_k$
defined in the paragraph after equation \eqref{decomposition}.

Because hermiticity of the quantum Laplace operator is the crucial property,
we define a quantized version of the action functional
\eqref{eq:ScalarAction} as
\begin{equation}
 S_k[\phi_k]:=\tr\big(\phi_k\hat{\Delta}_k(\phi_k)+V(\phi_k)\big)~,~~~\phi_k\in\End(E_k)~.
\end{equation}
As the functional $S_k$ lives on the finite dimensional space
$\End(E_k)$, the corresponding functional integral
\begin{equation}\label{eq:ScalarQuantumAction}
 Z=\int_{\End(E_k)} \CD[\phi_k]~e^{-S_k[\phi_k]}
\end{equation}
is a finite-dimensional integral and thus well-defined. This is what people
refer to as fuzzy quantum scalar field theory
\cite{Grosse:1995ar,Balachandran:2005ew,IuliuLazaroiu:2008pk}, and besides
providing a nice regularization procedure, using the quantized form
\eqref{eq:ScalarQuantumAction}, one can easily study the field theory
\eqref{eq:ScalarAction} numerically on a computer
\cite{GarciaFlores:2005xc,Panero:2006bx}.

\subsection{The ${\cal N}=(2,2)$ supersymmetric sigma model on $\C^{1|2}$}

Considering the supersymmetric sigma model with $(2,2)$ supersymmetries on the
superspace $\C^{1|2}$ is particular convenient as this space has the same
volume form as $\P^{1|2}$ on one of the standard patches, for which a bosonic
homogeneous coordinate, e.g.\ $z^0$, does not vanish. The reason for this is
that $\P^{1|2}$ is a Calabi-Yau supermanifold.

\paragraph{Calabi-Yau supermanifolds.} The spaces $\P^{n|n+1}$ come with a nowhere vanishing holomorphic volume form. Using the usual inhomogeneous coordinates $z_0^1,...,z_0^n,$ $\zeta_0^1,...,$ $\zeta_0^{n+1}$ on the patch $U_0:z^0\neq 0$ of $\P^{n|n+1}$, the superdeterminant in the Liouville measure \eqref{eq:LiouvilleMeasure} is just a constant, and thus
\begin{equation}
 \Omega^{n|n+1,0|0}_{U_0}:=\gamma~ z_0^1\wedge...\wedge d z_0^n
 d\zeta_0^1...d\zeta_0^{n+1}
\end{equation}
can be extended to a non-vanishing globally holomorphic volume form. Here,
$\gamma\in\FC^*$ is an arbitrary nonvanishing constant. (Recall that ${\rm
vol}(\P^{n|n+1})=0$, and therefore we cannot normalize by the space's natural
volume, as one would usually do.) The Liouville measure is then given by
$d\mu=\Omega^{n|n+1,0|0}\wedge \Omega^{0|0,n|n+1}$. It is evident that the
Berezinian super line bundle of these spaces is trivial. Such spaces are
referred to as {\em Calabi-Yau supermanifolds} in the literature. Note,
however, that Yau's theorem doesn't hold without restrictions in the
supercase, cf.\ \cite{Rocek:2004bi}. In particular, the spaces
$\P^{n|n+1}$ are not super Ricci-flat, i.e.\ the Ricci tensor
\begin{equation}
 R_{IJ}:=\derr{^2\log({\rm sdet(g)})}{Z^I\bZ^J}~,
\end{equation}
where $g$ is the super K{\"a}hler supermetric obtained from an arbitrary
K{\"a}hler form $\omega$, does not vanish. For our purposes, the existence of
$\Omega^{n|n+1,0|0}$, or equivalently, triviality of the Berezinian super line
bundle, will prove to be sufficient.

\

On $\FC^{1|2}$, we introduce the supercovariant derivatives
\begin{equation}
 D_{1,2}=\der{\zeta^{1,2}}\pm\zeta^{1,2}\der{z}~,~~~\bar{D}_{1,2}=\der{\bar{\zeta}^{1,2}}\pm\bar{\zeta}^{1,2}\der{\bz}
\end{equation}
as well as the generators for supersymmetry transformations
\begin{equation}
Q_{1,2}=\der{\zeta^{1,2}}\mp\zeta^{1,2}\der{z}~,~~~\bar{Q}_{1,2}=\der{\bar{\zeta}^{1,2}}\mp\bar{\zeta}^{1,2}\der{\bz}~.
\end{equation}
Note that the relation to the usual chiral notation is as follows:
\begin{equation}
 \zeta^+=\frac{1}{\sqrt{2}}(\zeta^1-\zeta^2)~,~~~\zeta^-=\frac{1}{\sqrt{2}}(\zeta^1+\zeta^2)~.
\end{equation}
The four basic superfields on $\FC^{1|2}$ are then given by (cf.\ e.g.\
\cite{Deligne:1999qp,Sevrin:1996jr})
\begin{equation}\label{def:ChiralSuperfields}
\begin{aligned}
 D_2\Phi_{\rm c}&=-D_1\Phi_{\rm c}~,~~~&\bar{D}_2\Phi_{\rm
 c}&=-\bar{D}_1\Phi_{\rm c}~,~~~~~&D_+\Phi_{\rm c}&=\bar{D}_+\Phi_{\rm c}=0\\
 D_2\Phi_{\rm ac}&=D_1\Phi_{\rm ac}~,~~~&\bar{D}_2\Phi_{\rm
 ac}&=\bar{D}_1\Phi_{\rm ac}~,~~~~~&D_-\Phi_{\rm ac}&=\bar{D}_-\Phi_{\rm
 ac}=0\\ D_2\Phi_{\rm tc}&=-D_1\Phi_{\rm tc}~,~~~&\bar{D}_2\Phi_{\rm
 tc}&=\bar{D}_1\Phi_{\rm tc}~,~~~~~&D_+\Phi_{\rm tc}&=\bar{D}_-\Phi_{\rm
 tc}=0\\ D_2\Phi_{\rm tac}&=D_1\Phi_{\rm tac}~,~~~&\bar{D}_2\Phi_{\rm
 tac}&=-\bar{D}_1\Phi_{\rm tac}~,~~~~~&D_-\Phi_{\rm tac}&=\bar{D}_+\Phi_{\rm
 tac}=0
\end{aligned} 
\end{equation}
corresponding to {\em chiral} (c), {\em anti-chiral} (ac), {\em twisted
chiral} (tc) and {\em twisted anti-chiral superfields} (tac), respectively. As
we are working in Euclidean space, the notion of reality is slightly more
subtle than in the Minkowski case. In particular, the ordinary complex
conjugate of a chiral superfield is not an antichiral superfield, and one has
to introduce a different real structure to obtain this result. However,
twisted chiral superfields are indeed complex conjugate to twisted anti-chiral
superfields and therefore we will restrict to them in most of the
following. For convenience, we introduce the chiral and anti-chiral
coordinates
\begin{equation}
\begin{aligned}
 z^+&:=z+\zeta^1\zeta^2=z+\zeta^+\zeta^-~,~~~&z^-&:=z-\zeta^1\zeta^2=z-\zeta^+\zeta^-~,\\
\bz^+&:=\bz+\bar{\zeta}^1\bar{\zeta}^2=\bz+\bar{\zeta}^+\bar{\zeta}^-~,~~~&\bz^-&:=\bz-\bar{\zeta}^1\bar{\zeta}^2=\bz-\bar{\zeta}^+\bar{\zeta}^-~,
\end{aligned}
\end{equation}
which satisfy $D_+z^-=D_-z^+=0$. From these fields, one can now construct an action using a real function $K(\Phi_{\rm c}^1,...,\Phi_{\rm c}^i,\Phi_{\rm c}^1,...,\Phi_{\rm ac}^i,\Phi_{\rm tc}^1,...,\Phi_{\rm tc}^j,\Phi_{\rm tac}^1,...,\Phi_{\rm tac}^j)=:K(\Phi)$ as follows:
\begin{equation}
 S=\int \d^2 z\d^2\zeta^1\d^2\zeta^2 K(\Phi)~.
\end{equation}
When interpreting the superfields $\Phi$ as maps from the worldsheet
$\C^{1|2}$ into a complex target manifold, one is led to regard the
superfunction $K$ as the K{\"a}hler potential of the target space if it only
depends on chiral and anti-chiral superfields. We have
\begin{equation}
 g_{ab}:=\frac{\dpar^2 K(\Phi)}{\dpar\Phi^a_{\rm c}\dpar\Phi^b_{\rm ac}}~,
\end{equation}
where $g_{ab}$ is the target space supermetric. If twisted chiral superfields
are included as well, there is an analogous relation to generalized complex
geometry.

One can furthermore add superpotential terms of the form
\begin{equation}
 \int \d^2z \d\zeta^-\d\zeta^+ W(\Phi_{\rm c})~,~~~\int \d^2z
 \d\bar{\zeta}^-\d\zeta^+ \hat{W}(\Phi_{\rm tc})~,
\end{equation}
which have to be accompanied by their complex conjugate. Here, $W$ and
$\hat{W}$ are polynomials in the chiral and twisted chiral superfields,
restricted by renormalizability of the theory.

To be concise, let us now restrict\footnote{It should be stressed, that more
general models could have been treated in principle.} to a specific model
which contains only twisted chiral superfields. (Recall that a sigma model
containing only twisted chiral superfields is dual to one containing only
untwisted ones). The superfield expansion of a twisted chiral superfield reads
as
\begin{equation*}
\begin{aligned}
 \Phi(z^+,\bz^-,\zeta^+,\bar{\zeta}^-)\ =\ &\phi(z^+,\bz^-)+\zeta^+\bar{\psi}^-(z^+,\bz^-)+\bar{\zeta}^-\psi^+(z^+,\bz^-)+\zeta^+\bar{\zeta}^-F(z^+,\bz^-)\\
\ =\ &\phi(z,\bz)+\zeta^+\zeta^-\dpar_z\phi(z,\bz)+\bar{\zeta}^+\bar{\zeta}^-\dpar_\bz\phi(z,\bz)+\zeta^+\zeta^-\bar{\zeta}^+\bar{\zeta}^-\dpar_z\dpar_\bz\phi(z,\bz)\\
&+\zeta^+\bar{\psi}^-(z,\bz)+\zeta^+\bar{\zeta}^+\bar{\zeta}^-\dpar_\bz\bar{\psi}(z,\bz)+\bar{\zeta}^-\psi^+(z,\bz)\\&+\bar{\zeta}^-\zeta^+\zeta^-\dpar_z\psi^+(z,\bz)+\zeta^+\bar{\zeta}^-F(z,\bz)~.
\end{aligned} 
\end{equation*}
Putting
\begin{equation}
 K(\Phi)=\bar{\Phi}_{\rm tc}\Phi_{\rm tc}\mbox{~~~and~~~}\hat{W}(\Phi_{\rm
 tc})=m\Phi_{\rm tc}^2+\lambda \Phi_{\rm tc}^3~,
\end{equation}
we arrive at a sigma model with the component action
\begin{equation}
\begin{aligned}
 S=\int d^2z
~\Big(&\bar{\phi}\dpar_z\dpar_\bz\phi+\dpar_\bz\bar{\phi}\dpar_z\phi+(\dpar_z\dpar_\bz\bar{\phi})\phi+F\bar{F}\\
&-\psi^-\dpar_z\psi^++\bar{\psi}^+\dpar_\bz\bar{\psi}^--(\dpar_\bz\psi^-)\psi^++(\dpar_z\bar{\psi}^+)\bar{\psi}^-\\
&+2m^2(\phi
F-\bar{\psi}^-\psi^+)+3\lambda(\phi^2F-\phi\bar{\psi}^-\psi^+)+c.c.\Big)~,
\end{aligned} 
\end{equation}
where all fields depend only on $z$ (non-holomorphically, in general). After
integrating out the auxiliary fields and integrating by parts, we arrive at
the final form of the action
\begin{equation}\label{SUSYaction}
\begin{aligned}
 S=\int d^2 z~\Big(&\bar{\phi}\dpar_z\dpar_\bz\phi
-\psi^-\dpar_z\psi^++\bar{\psi}^+\dpar_\bz\bar{\psi}^--(\dpar_\bz\psi^-)\psi^++(\dpar_z\bar{\psi}^+)\bar{\psi}^-\\
&+3|2m\phi+3\lambda\phi^2|^2\Big)~.
\end{aligned}
\end{equation}

\subsection{Regularization with Berezin-quantized $\P^{1|2}$}

To regularize the theory \eqref{SUSYaction}, we would like to obtain a
supersymmetric theory on $\P^{1|2}$, which, upon decompactification (or,
equivalently, taking out a (super)point) turns into the supersymmetric
sigma-model on $\FC^{1|2}$. We can then translate the theory from $\P^{1|2}$
to Berezin-quantized $\P^{1|2}$ to obtain a finite quantum field theory.

Two issues remain to be clarified. The first one concerns the definition of chiral and twisted chiral superfields on $\P^{1|2}$ and the relation with supersymmetry transformations, while the second one is the integration over chiral and twisted chiral superspace.

As the space $\P^{1|2}$ is group theoretically given by the coset space
$U(2|2)/(U(1|0)\times U(1|2))$, its isometry group\footnote{Here, we choose to
use the full unitary supergroup to avoid discussing the projective subgroup
$PSU(2|2)$.} is $U(2|2)$. We will work at the level of the algebra of
generators $u(2|2)$, and we will use the following Hermitian generators:
\begin{equation}
 (\sigma^{IJ})_{AB}=\varphi_{IJ}\delta_{IA}\delta_{JB}+\varphi_{JI}\delta_{IB}\delta_{JA}~~\mbox{and}~~~(\rho^{IJ})_{AB}=i\varphi_{IJ}\delta_{IA}\delta_{JB}-i\varphi_{JI}\delta_{IB}\delta_{JA},
\end{equation}
where $I,J,A,B\in 0,...,3$ and $\varphi_{IJ}=e^{\pi i/2 \tilde{I}-\pi i/2
\tilde{J}}$ is a phase factor necessary to guarantee that our norm of vectors
in $\FC^{2|2}$ is invariant. In particular, $\sigma^{IJ}$ and $\rho^{IJ}$
generate space-time rotations for $I\leq1,J\leq1$, R-symmetry rotations for
$I\geq2,J\geq2$ and supersymmetry transformations in all other cases.

The representation $R$ of these generators acting on superfunctions on
$\FC^{2|2}$ is given by the differential operators
\begin{equation}
R(\sigma^{IJ})=Z^I\sigma^{IJ}\der{Z^J}-\bZ^I\sigma^{IJ}\der{\bZ^J}~~~\mbox{and}~~~R(\rho^{IJ})=Z^I\rho^{IJ}\der{Z^J}+\bZ^I\rho^{IJ}\der{\bZ^J}~,
\end{equation}
where $(Z^I)=(z^0,z^1,\zeta^1,\zeta^2)$ are the coordinates on
$\FC^{2|2}$. Note that $|Z|^2$ is invariant as expected. To obtain the
corresponding action on $\P^{1|2}$, we have these symmetries act on a certain
patch $U$ on the inhomogeneous coordinates $Z^I_0$. Consider again the patch
$U_0$ for which $z^0\neq 0$, then we have in addition to the generators
\begin{equation*}
 R_0(\sigma^{IJ})=Z_0^I\sigma^{IJ}\der{Z_0^J}-\bZ_0^I\sigma^{IJ}\der{\bZ_0^J}~~~\mbox{and}~~~R_0(\rho^{IJ})=Z^I_0\rho^{IJ}\der{Z^J_0}+\bZ^I_0\rho^{IJ}\der{\bZ^J_0}~~~,~I,J\geq1
\end{equation*}
the generators
\begin{equation*}
\begin{aligned}
R_0(\sigma^{00})=0~,~~
R_0(\sigma^{0I})=\der{Z^I_0}-Z^I_0\CE-\der{\bZ^I_0}+\bZ^I_0\bar{\CE}~,\\
R_0(\rho^{0I})=i\left(\der{Z^I_0}+Z^I_0\CE+\der{\bZ^I_0}+\bZ^I_0\bar{\CE}\right)~~~~~~~~~
\end{aligned}
\end{equation*}
for $i\leq1$ and
$\CE:=z^1_0\dpar_{z^1_0}+\zeta^1_0\dpar_{\zeta^1_0}+\zeta^2_0\dpar_{\zeta^2_0}$. Note
that the expression
$|Z_0|^2:=1+z\bz+i\zeta^1\bar{\zeta}^1+i\zeta^2\bar{\zeta}^2$ is only
invariant under transformations $R_0(\sigma^{IJ})$ with $I,J\geq 1$. When
decompactifying $\P^{1|2}$ to $\FC^{1|2}$, the Euler operators $\CE$ vanish,
and we can thus identify the differential operators $D_1$ and $D_2$ with the
generators according to
\begin{equation}
\begin{aligned}
 D_1=D_1^{R_0}=\tfrac{1}{2}(R_0(\sigma^{02})-i
 R_0(\rho^{02})+R_0(\sigma^{21})-i R_0(\rho^{21}))~,\\
 D_2=D_2^{R_0}=\tfrac{1}{2}(R_0(\sigma^{03})-i
 R_0(\rho^{03})-R_0(\sigma^{31})+i R_0(\rho^{31}))~.
\end{aligned}
\end{equation}
Using the corresponding differential operators $D_{1,2}^R$ in the
representation $R$, we have an action on monomials in the homogeneous
coordinates, which preserves their bi-degree. Recall that superfunctions on
$\P^{1|2}$ are written in terms of basis superfunctions
\begin{equation}
 \frac{Z^{I_1}...Z^{I_k}\bZ^{J_1}...\bZ^{J_k}}{|Z|^{2k}}~,
\end{equation}
and the action of $D_{1,2}^R$ on these superfunctions is given as the action
of the differential operators in coordinates of $\FC^{2|2}$ on the
numerator. (The denominator is invariant under $u(2|2)$-transformations.) This
allows us to define all the chiral superfields as above in
\eqref{def:ChiralSuperfields}. Note that a superfield of any of the possible
chiralities will transform into a non-chiral superfield under arbitrary
$u(2|2)$ supersymmetry transformations, as $D_{1,2}^R$ does not anticommute
with general supersymmetry transformations. However, the number of independent
component fields remains evidently the same and is merely reshuffled in the
field expansion. We will come back to this point later.

The second issue is the integration over chiral and anti-chiral superspace to
allow for the inclusion of a non-trivial superpotential. As the only invariant
measure available is the full integral over superspace, we have to insert a
superfunction, which takes care of the antichiral part:
\begin{equation}
 \int d^2z d\zeta^+d\bar{\zeta}^-~\rightarrow~\int d\mu([Z])
 \frac{\bar{\zeta}^+\zeta^-}{|Z|^2}~,
\end{equation}
where $d \mu([Z])$ is again the super Liouville measure on $\P^{1|2}$. Note
that indeed $\frac{\bar{\zeta}^+\zeta^-}{|Z|^2}\in
\CC^\infty(\P^{1|2})$. After integrating out the auxiliary fields, the factor
$\frac{1}{|Z|^2}$ will produce a factor of $\frac{1}{|z|^4}$ in front of
potential terms, the usual Liouville measure on $\P^1$. This will produce the
correct planar limit, after decompactifying $\P^1$ to $\C^1$.

To regularize this model on $\P^{1|2}$ by Berezin-quantizing the worldsheet as
$(\P^{1|2},E:=O(k))$, we need to translate all the above machinery to the
quantum situation. First, superfields are now elements of $\End_k$, and this
space is spanned by the operators
\begin{equation}
 \hat{A}^\dagger_{I_1}...\hat{A}^\dagger_{I_k}|0\rangle\langle
 0|\hat{A}_{J_1}...\hat{A}_{J_k}~,
\end{equation}
cf.\ section 5. The $u(2|2)$ invariant integral is given by the supertrace
\begin{equation}
 \int_{\P^{1|2}} d\mu(z)\sigma(\hat{f})=\frac{{\rm
 vol}'(\P^{1|2})}{b^0_k+b^1_k}{\rm str}(\hat{f})
\end{equation}
and the representation $\hat{R}$ of the generators $\sigma^{IJ}$ and
$\rho^{IJ}$ on $\End_k$ is the usual Schwinger representation
\begin{equation}
 \hat{R}(\sigma^{IJ})(\hat{f})=\lsc \hat{A}^\dagger_A
 \sigma^{IJ}_{AB}\hat{A}_B,\hat{f}\rsc~,~~~\hat{R}(\rho^{IJ})(\hat{f})=\lsc\hat{A}^\dagger_A
 \rho^{IJ}_{AB}\hat{A}_B,\hat{f}\rsc~.
\end{equation}
While the definition of twisted chiral and twisted anti-chiral superfields in
this manner goes over into ordinary twisted chiral and twisted anti-chiral
superfields upon decompactification, one might argue that it is still too
restrictive. Having in mind that $D_1$ and $D_2$ act only holomorphically on
the fields, one could restrict the actions of $\hat{R}(\sigma^{IJ})$ in the
definition of twisted chiral superfields to their left-actions, which amounts
to a holomorphic action on the corresponding fields. This point is quite
subtle and requires certainly further scrutiny.

The integral over chiral superspace can now be performed in two different
ways. Either, we multiply the operator to be integrated over chiral superspace
with the operator corresponding to the superfunction
$\frac{\bar{\zeta}^+\zeta^-}{|Z|^2}$ and integrate via the supertrace on
$\End_k$, or we add the corresponding creation and annihilation operators to
all superfunctions by insertion and integrate by taking the supertrace over
$\End_{k+1}$. Here, we will choose to work with the former procedure. Putting
everything together, we have the following action:
\begin{equation}\label{regaction}
 S=\frac{{\rm vol}'(\P^{1|2})}{b^0_k+b^1_k}~{\rm
str}\Big(\hat{\Phi}^\dagger_{\rm tc}\hat{\Phi}_{\rm tc}+\hat{\Psi}
\hat{W}(\hat{\Phi}_{\rm
tc})+\hat{\Psi}^\dagger\hat{W}^\dagger(\hat{\Phi}^\dagger_{\rm tc}) \Big)~,
\end{equation}
where $\hat{W}(\hat{\Phi}_{\rm ch})$ is again a polynomial in its argument
$\hat{\Phi}_{\rm ch}$ and
\begin{equation}
 \hat{\Psi}:=\alpha^\dagger_+\hat{A}^\dagger_{I_1}...\hat{A}^\dagger_{I_{k-1}}|0\rangle\langle
 0|\hat{A}_{I_{k-1}}...\hat{A}_{I_1}\alpha_-~.
\end{equation}
The superfunctional integral has now to be taken over all operators
corresponding to twisted chiral fields
\begin{equation}
 Z\ =\ \int \CD \Phi_{\rm tc} \exp(-S)~;
\end{equation}
it is a finite integral and thus provides a regularization of the ${\cal
N}=(2,2)$ supersymmetric sigma model in two dimensions in the usual sense of
fuzzy geometry.

\paragraph{Remarks.} The original sigma-model on $\FC^{1|2}$ was invariant under 4 (real) supercharges: $Q_\pm, \bar{Q}_\pm$. The algebra of isometries of $\P^{1|2}$ contains, however, 8 odd generators. This shows up in the fact that the definition of a twisted chiral superfield is not invariant\footnote{This is clear from group theoretic considerations.} under half of the $u(2|2)$ generators. Without superpotential term, the global symmetry group of the action \eqref{regaction} is indeed $U(2|2)$, and the supersymmetry transformations modifying twisted chiral superfields merely reshuffle the component fields. This invariance is easily seen as the D-term ${\rm str}(\hat{\Phi}^\dagger_{\rm tc}\hat{\Phi}_{\rm tc})$ is evidently invariant under transformations $\hat{\Phi}_{\rm tc}\rightarrow \hat{U}\hat{\Phi}_{\rm tc}\hat{U}^\dagger$. A superpotential term, however, breaks the supersymmetry of the model down to the same as the one on $\FC^{1|2}$, which we set out to regularize in the first place, and this was in fact to be expected.

\subsection{Comments on the topological twist}

In a more general context, the above mentioned sigma model can be defined on an arbitrary Riemann surface with canonical bundle $K$. The associated super Riemann surface is a split supermanifold which is the total space of the (real) rank 4 vector bundle
\begin{equation}
 (\Pi K^{1/2}\oplus\Pi\bar{K}^{1/2})\oplus(\Pi
 K^{1/2}\oplus\Pi\bar{K}^{1/2})~. \label{topvb}
\end{equation}
In the language of \cite{Manin:1988ds} and section 2, this is a superspace
$(X,\cA)$ such that $X_\red$ is a Riemann surface and $\hat{\cA}$ isomorphic
to the sheaf of supersections of the vector bundle $(\ref{topvb})$ with $\cA$
globally isomorphic to $\wedge^n_{\cA_\red} \hat{\cA}$.  In our example,
$K=O(2)$ and $\zeta^+,\bar{\zeta}^+$ are supersections of the first two
super line bundles, while $\zeta^-,\bar{\zeta}^-$ are supersections of the
second two. Applying a topological twist (see e.g.\ \cite{Deligne:1999qp}), we
deform this geometry to
\begin{equation}
 (\Pi O\oplus\Pi K)\oplus(\Pi O\oplus\Pi\bar{K})~.
\end{equation}
In our example, the resulting space would be the weighted superprojective
space\linebreak $W\P^{1|2}(1,1|0,2)$, which is the total space of the vector
bundle $O\oplus\Pi O(2)$ over $\P^1$.

While on flat space, this twist corresponds to a mere rewriting, on curved
space, the twist allows for defining supersymmetric models on non-spin
manifolds and avoids the introduction of spinors altogether. In particular,
the Gra\ss mann coordinates parametrizing the trivial super line bundle give
rise to supercharges which carry Lorentz spin 0 and are thus invariant under
space-time rotations. This guarantees the preservation of a certain amount of
supersymmetry.

The definition of chiral and twisted chiral fields on this geometry proceeds
as before, and following the procedure of the untwisted case, one eventually
arrives at a topologically twisted sigma model on quantized
$W\P^{1|2}(1,1|0,2)$.

\section{Summary and directions for further research}

In this paper, we defined generalized Berezin and Berezin-Toeplitz quantization of Hodge supermanifolds. A prerequisite for this quantization was the given extension of the Rawnsley coherent states to the case of supermanifolds. Explicitly, we constructed the quantization of both affine and projective superspaces. Eventually, we showed how one can employ such quantized supermanifolds as supersymmetry-preserving regulators of quantum field theories; we proposed definitions of ordinary and twisted $\cN=(2,2)$ supersymmetric sigma models on the compactified superspace $\P^{1|2}$.

Taking our results as a starting point, one has a number of potentially interesting directions for future research. Clearly, it would be desirable to expose more supersymmetric field theories admitting a regularization by Berezin-quantized Hodge supermanifolds. One is evidently restricted to such theories which allow for a superfield formulation. However it is unclear, whether one is limited to using the quantizations of Calabi-Yau supermanifolds in regularizing supersymmetric field theories on flat superspace. Moreover, an extension to supersymmetric gauge theories is desirable, having in mind the ultimate aim of the minimal supersymmetric standard model regularized on a fuzzy superspace. Also, one would expect that the topological twist plays a crucial role in regularizing supersymmetric field theories using more general quantized Hodge supermanifolds as it allows for working without spinors.

A natural question with respect to the nonlinear sigma models regularized above would be whether mirror symmetry holds after regularization. This would require a more general analysis of ${\cal N}=(2,2)$ supersymmetric nonlinear sigma models on Calabi-Yau manifolds but this ``fuzzy mirror symmetry'' would be useful in the associated ${\cal N}=2$ superconformal algebra calculations and thus it might help with numerical studies of mirror symmetry.

Numerical studies\footnote{For recent work in this direction, see \cite{Bietenholz:2008ny}.} of the models proposed above and their generalizations can be readily performed, and the behavior of the regulated models should be compared to the conventional knowledge of supersymmetric field theories. Note also that here, one is analyzing a supermatrix model, and the application of matrix model techniques to these regulated supersymmetric field theories in the spirit of \cite{O'Connor:2007ea} might yield more interesting results than in the non-supersymmetric case. 

More formally, it seems to be a mere technicality to extend the relation between geometric quantization and formal deformation quantization using Berezin-Toeplitz quantization to the case of supermanifolds. Eventually, one might wish to extend the known relationship between quantizable Hermitian symmetric spaces and the Toeplitz quantization procedure \cite{Tuynman:1987jc} to the case of supermanifolds.

\acknowledgments
CS would like to thank Christoph Sachse for discussions on supermathematics and in particular Denjoe O'Connor for many discussions on fuzzy supersymmetric field theories in the past. DM is supported by an IRCSET (Irish Research Council for Science, Engineering
and Technology) postgraduate research scholarship. CS is supported by an IRCSET postdoctoral fellowship.


\begin{thebibliography}{10}

\bibitem{Woodhouse:1992de}
N.~M.~J.~Woodhouse,
{\em {Geometric quantization},}
Oxford mathematical monographs, New York, USA: Clarendon (1992).

\bibitem{Grosse:1995ar}
H.~Grosse, C.~Klimcik, and P.~Presnajder,
{\em Towards finite quantum field theory in noncommutative geometry,}
Int. J. Theor. Phys. {\bf 35} (1996) 231 [{\tt hep-th/9505175}].

\bibitem{Myers:1999ps}
R.~C.~Myers,
{\em Dielectric-branes,}
JHEP {\bf 12} (1999) 022 [{\tt hep-th/9910053}].

\bibitem{Seiberg:1999vs}
N.~Seiberg and E.~Witten,
{\em String theory and noncommutative geometry,}
JHEP {\bf 09} (1999) 032 [{\tt hep-th/9908142}].

\bibitem{Kostant-1970aa}
B.~Kostant,
{\em Quantization and unitary representations,}
in: ``Lecture Notes in Mathematics III'', p.87, Springer (1970).

\bibitem{Kostant:1975qe}
B.~Kostant,
{\em {Graded manifolds, graded Lie theory, and prequantization},}
in: ``Bonn 1975, Proceedings, Differential geometrical methods in mathematical
  physics,'' Berlin 1977, 177.

\bibitem{Souriau-1970aa}
J.-M.~Souriau,
{\em Structure des systemes dynamiques,}
Dunod, Paris (1969).

\bibitem{Tuynman:1992zm}
G.~M.~Tuynman,
{\em {Geometric quantization of the BRST charge},}
Commun. Math. Phys. {\bf 150} (1992) 237.

\bibitem{MR1032867}
C.~LeBrun, Y.~S.~Poon, and R.~O.~Wells, Jr.,
{\em Projective embeddings of complex supermanifolds,}
Comm. Math. Phys. {\bf 126} (1990) 433.

\bibitem{IuliuLazaroiu:2008pk}
C.~Iuliu-Lazaroiu, D.~McNamee, and C.~Saemann,
{\em {Generalized Berezin quantization, Bergman metrics and fuzzy Laplacians},}
JHEP {\bf 09} (2008) 059 [{\tt 0804.4555 [hep-th]}].

\bibitem{ElGradechi:1993gq}
A.~M.~El~Gradechi,
{\em {On the supersymplectic homogeneous superspace underlying the OSp(1/2)
  coherent states},}
J. Math. Phys. {\bf 34} (1993) 5951 [{\tt hep-th/9301132}];
A.~M.~El~Gradechi and L.~M.~Nieto,
{\em {Supercoherent states, superKahler geometry and geometric quantization},}
Commun. Math. Phys. {\bf 175} (1996) 521 [{\tt hep-th/9403109}].

\bibitem{Grosse:1995pr}
H.~Grosse, C.~Klimcik, and P.~Presnajder,
{\em Field theory on a supersymmetric lattice,}
Commun. Math. Phys. {\bf 185} (1997) 155 [{\tt hep-th/9507074}].

\bibitem{Murray:2006pi}
S.~Murray and C.~Saemann,
{\em Quantization of flag manifolds and their supersymmetric extensions,}
Adv. Theor. Math. Phys. {\bf 12} (2008) 641 [{\tt hep-th/0611328}].

\bibitem{Perelomov:1986tf}
A.~M.~Perelomov, {\em Generalized coherent states and their applications,}
Springer, Berlin (1986).

\bibitem{Berezin:1974du}
F.~A.~Berezin,
{\em General concept of quantization,}
Commun. Math. Phys. {\bf 40} (1975) 153.

\bibitem{Tuynman:1987jc}
G.~M.~Tuynman,
{\em Quantization: Towards a comparison between methods,}
J. Math. Phys. {\bf 28} (1987) 2829.

\bibitem{MaMarinescu}
X.~Ma, G.~Marinescu,
{\em Toeplitz operators on symplectic manifolds,}
J. Geom. Anal. {\bf 18}, 2, (2008), 565-611.

\bibitem{Manin:1988ds}
Y.~I.~Manin,
{\em Gauge field theory and complex geometry,}
Grundlehren der mathematischen Wissenschaften, 289, Springer (1988).

\bibitem{Varadarajan:2004yz}
V.~S.~Varadarajan,
{\em {Supersymmetry for mathematicians: An introduction},}
New York, USA: Courant Inst. Math. Sci. (2004).

\bibitem{Haske-1987aa}
C.~Haske and R.~O.~Wells,
{\em Serre duality on complex supermanifolds,}
Duke Math. J. {\bf 54} (1987) 493.

\bibitem{Rawnsley:1976gb}
J.~Rawnsley,
{\em Coherent states and K{\"a}hler manifolds,}
Quat. J. Math. Oxford {\bf 28} (1977) 403.

\bibitem{Rawnsley:1990tj}
J.~Rawnsley, M.~Cahen, and S.~Gutt,
{\em Quantization of K{\"a}hler manifolds. I: Geometric interpretation of
  Berezin's quantization,}
J. Geom. Phys. {\bf 7} (1990)~45.

\bibitem{Rawnsley-1993aa}
J.~Rawnsley, M.~Cahen, and S.~Gutt,
{\em Quantization of K{\"a}hler manifolds II,}
Trans. Amer. Math. Soc. {\bf 337} (1993)~73.

\bibitem{Donaldson:2001aa}
S.~K.~Donaldson,
{\em Scalar curvature and projective embeddings, I,}
J. Diff. Geom. {\bf 59} (2001) 479.

\bibitem{Catenacci:arXiv0707.4246}
R.~Catenacci, M.~Debernardi, P.~A.~Grassi, and D.~Matessi,
{\em Balanced superprojective varieties,}
{\tt 0707.4246 [math-ph]}.

\bibitem{Tian-1990}
G.~Tian,
{\em On a set of polarised K{\"a}hler metrics on algebraic manifolds,}
J. Diff. Geom. {\bf 32} (1990)~99.

\bibitem{Schlichenmaier-1996aa}
M.~Schlichenmaier,
{\em Berezin-Toeplitz quantization of compact K{\"a}hler manifolds,}
in: ``Quantization, coherent states and Poisson structures'' (Bialowiecza
  1995), 101, PWN Warsaw, 1998, Mannheimer Manuskripte Nr. 203 [{\tt
  q-alg/9601016}].

\bibitem{Schlichenmaier-1999aa}
M.~Schlichenmaier,
{\em Berezin-Toeplitz quantization and Berezin symbols for arbitrary compact
  Kaehler manifolds,}
{\tt math.QA/9902066}.

\bibitem{Schlichenmaier-1999bb}
M.~Schlichenmaier,
{\em Deformation quantization of compact K{\"a}hler manifolds by
  Berezin-Toeplitz quantization,}
Conf\'erence Mosh\'e Flato 1999, Vol. II (Dijon) 289, Math. Phys. Stud. {\bf 22},
  Kluwer, Dordrecht, 2000 [{\tt math.QA/9910137}].

\bibitem{borthwick}
D.~Borthwick, S.~Klimek, A.~Lesniewski, M.~Rinaldi,
{\em Super Toeplitz operators and non-perturbative deformation quantization of supermanifolds,}
Commun. Math. Phys. {\bf 153} (1993), 49-76.


\bibitem{bordemanndefquant}
M.~Bordemann,
{\em The deformation quantization of certain super-Poisson brackets and BRST cohomology,}
Conf\'erence Mosh\'e Flato 1999, Vol. II (Dijon) 289, Math. Phys. Stud. {\bf 22},
  Kluwer, Dordrecht, 2000, [{\tt math.QA/0003218}].

\bibitem{Balachandran:2002jf}
A.~P.~Balachandran, S.~Kurkcuoglu, and E.~Rojas,
{\em The star product on the fuzzy supersphere,}
JHEP {\bf 07} (2002) 056 [{\tt hep-th/0204170}].

\bibitem{Sachse:2008aa}
C.~Sachse,
{\em A categorical formulation of superalgebra and supergeometry,}
{\tt 0802.4067 [math.AG]}.

\bibitem{Ivanov:2003qq}
E.~Ivanov, L.~Mezincescu, and P.~K.~Townsend,
{\em Fuzzy $CP^{(n|m)}$ as a quantum superspace,}
{\tt hep-th/0311159}.

\bibitem{bordemanncpn1}
M.~Bordemann, M.~Brischle, C.~Emmrich, S.~Waldmann,
{\em Phase Space Reduction
for Star Products: An Explicit Construction for $\C P^n$,}
Lett. Math. Phys. {\bf 36} (1996), 357-371, [{\tt q-alg/9503004}].

\bibitem{bordemanncpn2}
M.~Bordemann, M.~Brischle, C.~Emmrich, S.~Waldmann,
{\em Subalgebras with converging star products in deformation quantization: An
algebraic construction for $\C P^n$,}
J. Math. Phys. {\bf 37} (1996), 6311-6323, [{\tt q-alg/9512019}].

\bibitem{Saemann:2006gf}
C.~Saemann,
{\em Fuzzy toric geometries,}
JHEP {\bf 02} (2008) 111 [{\tt hep-th/0612173}].

\bibitem{Klimcik:1999pr}
C.~Klimcik,
{\em An extended fuzzy supersphere and twisted chiral superfields,}
Commun. Math. Phys. {\bf 206} (1999) 587 [{\tt hep-th/9903202}].

\bibitem{Hubsch:1998ps}
T.~Hubsch,
{\em {Haploid (2,2)-superfields in 2-dimensional spacetime},}
Nucl. Phys. B {\bf 555} (1999) 567 [{\tt hep-th/9901038}].

\bibitem{Balachandran:2005ew}
A.~P.~Balachandran, S.~Kurkcuoglu, and S.~Vaidya,
{\em Lectures on fuzzy and fuzzy SUSY physics,}
{\tt hep-th/0511114}.

\bibitem{GarciaFlores:2005xc}
F.~Garcia~Flores, D.~O'Connor, and X.~Martin,
{\em Simulating the scalar field on the fuzzy sphere,}
PoS LAT {\bf 2005} (2005) 262 [{\tt hep-lat/0601012}].

\bibitem{Panero:2006bx}
M.~Panero,
{\em Numerical simulations of a non-commutative theory: The scalar model on the
  fuzzy sphere,}
JHEP {\bf 05} (2007) 082 [{\tt hep-th/0608202}].

\bibitem{Rocek:2004bi}
M.~Rocek and N.~Wadhwa,
{\em On Calabi-Yau supermanifolds,}
Adv. Theor. Math. Phys. {\bf 9} (2005) 315 [{\tt hep-th/0408188}];
M.~Rocek and N.~Wadhwa,
{\em On Calabi-Yau supermanifolds. II,}
{\tt hep-th/0410081}.

\bibitem{Deligne:1999qp}
P.~Deligne, P.~Etingof, D.~Freed, L.~Jeffrey, D.~Kazhdan, J.~Morgan,
  D.~Morrison, and E.~Witten,
{\em Quantum fields and strings: A course for mathematicians,}
AMS, Providence (1999).

\bibitem{Sevrin:1996jr}
A.~Sevrin and J.~Troost,
{\em {Off-shell formulation of N = 2 non-linear sigma-models},}
Nucl. Phys. B {\bf 492} (1997) 623 [{\tt hep-th/9610102}].

\bibitem{Bietenholz:2008ny}
W.~Bietenholz,
{\em {Simulations of a supersymmetry inspired model on a fuzzy sphere},}
Contributed to 25th International Symposium on Lattice Field Theory,
  Regensburg, Germany, 30 Jul - 4 Aug 2007 [{\tt 0808.2387 [hep-th]}].

\bibitem{O'Connor:2007ea}
D.~O'Connor and C.~Saemann,
{\em Fuzzy scalar field theory as a multitrace matrix model,}
JHEP {\bf 08} (2007) 066 [{\tt 0706.2493 [hep-th]}].

\end{thebibliography}
\end{document}